\documentclass[acmsmall]{acmart}
\usepackage{booktabs}
\usepackage{multirow}
\usepackage{threeparttable}
\usepackage[many]{tcolorbox} 
\usepackage{fontawesome5} 
\usepackage{colortbl}

\AtBeginDocument{%
  }
\newtcolorbox{boxK}{
    top=2pt,
    bottom=2pt,
    left=2pt,
    right=2pt,
    boxrule = 0pt,
    toprule = 0pt, 
}

\setcopyright{acmlicensed}
\acmDOI{10.1145/3660778}
\acmYear{2024}
\copyrightyear{2024}
\acmSubmissionID{fse24main-p118-p}
\acmJournal{PACMSE}
\acmVolume{1}
\acmNumber{FSE}
\acmArticle{71}
\acmMonth{7}
\received{2023-09-28}
\received[accepted]{2024-04-16}

\begin{document}

\title{Exploring and Unleashing the Power of Large Language Models in Automated Code Translation}

\author{Zhen Yang}
\orcid{0000-0003-0670-4538}
\affiliation{%
  \institution{Shandong University}
  \city{Qingdao}
  \country{China}
}
\email{zhenyang@sdu.edu.cn}

\author{Fang Liu}\authornote{Corresponding Author}
\orcid{0000-0002-3905-8133}
\affiliation{%
  \institution{Beihang University}
  \city{Beijing}
  \country{China}
}
\email{fangliu@buaa.edu.cn}

\author{Zhongxing Yu}\authornotemark[1]
\orcid{0000-0003-3718-8476}
\affiliation{%
  \institution{Shandong University}
  \city{Qingdao}
  \country{China}
}
\email{zhongxing.yu@sdu.edu.cn}

\author{Jacky Wai Keung}
\orcid{0000-0002-3803-9600}
\affiliation{%
  \institution{City University of Hong Kong}
  \city{Hong Kong}
  \country{China}
}
\email{Jacky.Keung@cityu.edu.hk}

\author{Jia Li\faMars}
\orcid{0000-0002-5579-8852}
\affiliation{%
  \institution{Peking University}
  \city{Beijing}
  \country{China}
}
\email{lijia@stu.pku.edu.cn}

\author{Shuo Liu}
\orcid{0000-0002-8877-3678}
\email{sliu273-c@my.cityu.edu.hk}
\author{Yifan Hong}
\orcid{0009-0004-9832-8879}
\email{yifanhong7-c@my.cityu.edu.hk}
\author{Xiaoxue Ma}
\orcid{0000-0002-5476-6074}
\email{xiaoxuema3-c@my.cityu.edu.hk}
\affiliation{%
  \institution{City University of Hong Kong}
  \city{Hong Kong}
  \country{China}
}

\author{Zhi Jin}
\orcid{0000-0003-1087-226X}
\email{zhijin@pku.edu.cn}
\author{Ge Li}
\orcid{0000-0002-5828-0186}
\email{lige@pku.edu.cn}
\affiliation{%
  \institution{Peking University}
  \city{Beijing}
  \country{China}
}

\renewcommand{\shortauthors}{Zhen Yang, Fang Liu, Zhongxing Yu, Jacky Wai Keung, et al.}

\begin{abstract}
Code translation tools, namely transpilers, are developed for automatic source-to-source translation. Latest learning-based transpilers have shown impressive enhancement against rule-based counterparts in both translation accuracy and readability, owing to their task-specific pre-training on extensive monolingual corpora. Nevertheless, their current performance still remains unsatisfactory for practical deployment, and the associated training resources are also prohibitively expensive. Large Language Models (LLMs), pre-trained on huge amounts of human-written code/text, have shown remarkable performance in many code intelligence tasks due to their powerful generality, even without task-specific re-training/fine-tuning. Thus, LLMs can potentially circumvent the above limitations, but they have not been exhaustively explored yet. This paper investigates diverse LLMs and learning-based transpilers for automated code translation tasks, finding that: although certain LLMs have outperformed current transpilers, they still have some accuracy issues, where most of the failures are induced by a lack of comprehension of source programs (38.51\%), missing clear instructions on I/O types in translation (14.94\%), and ignoring discrepancies between source and target programs (41.38\%).

Enlightened by the above findings, we further propose \textbf{UniTrans}, a \textbf{Uni}fied code \textbf{Trans}lation framework, applicable to various LLMs, for unleashing their power in this field. 
Specifically, \textbf{UniTrans} first crafts a series of test cases for target programs with the assistance of source programs. Next, it harnesses the above auto-generated test cases to augment the code translation and then evaluate their correctness via execution. Afterward, \textbf{UniTrans} further (iteratively) repairs incorrectly translated programs prompted by test case execution results.
Extensive experiments are conducted on six settings of translation datasets between Python, Java, and C++. Three recent LLMs of diverse sizes, including GPT-3.5 and LLaMA-13B/7B, are tested with \textbf{UniTrans}, and all achieve substantial improvements in terms of computational accuracy and exact match accuracy among almost all translation settings, showing the universal effectiveness of \textbf{UniTrans} in practice.
\end{abstract}

\ccsdesc[500]{Software and its engineering~Genetic programming}

\keywords{Automated Code Translation, Large Language Models, Transformer}

\maketitle

\section{Introduction}
With the advancement and prosperity of software, various Programming Languages (PLs) were invented to deal with diverse development scenarios and needs, such as desktop applications, websites, and mobile applications. More and more software developed in one PL also has the necessity to be ported to other languages to satisfy the extension of business platforms \cite{10.1109/ICSE48619.2023.00072,10.1145/1806799.1806848,10.1109/ASE.2015.74}. Therefore, different transpilers \cite{allamanis2018survey, karaivanov2014phrase,nguyen2016mapping, roziere2020unsupervised, roziere2021leveraging} aiming at source-to-source translations emerged during the last several decades to improve the codebase migration efficiency. 

Traditional rule-based transpilers require not only expertise in both source and target languages but also considerable effort and time-cost in rule design \cite{roziere2020unsupervised}. 
For example, the Commonwealth Bank of Australia invested approximately $\$$750 million and dedicated five years to migrate their platform from COBOL to Java \cite{10.1145/1806799.1806831}. Meanwhile, the translation performance in readability and correctness are also poor \cite{roziere2020unsupervised}. Therefore, a series of learning-based transpilers were proposed to improve the translation efficiency and efficacy with Neural Machine Translation (NMT) methods \cite{roziere2020unsupervised,roziere2021leveraging,szafraniec2022code,chen2018tree}, which
normally leverage diverse task-specific pre-training on huge amounts of monolingual corpora.
Previous studies have demonstrated the impressive improvement of these learning-based transpilers, but their current performance is still unsatisfactory for practical deployment, and the training resources are also unaffordable.
For example, TransCoder \cite{roziere2020unsupervised}, one of the state-of-the-art learning-based transpilers, was trained on 32 V100 GPUs for 12 days, whereas its translation performance among various mainstream PL pairs all cannot reach 50\% in terms of computational accuracy according to our empirical results.
Recent studies reveal that Large Language Models (LLMs), pre-trained on billions of text/code tokens, bypass the need for re-training/fine-tuning but demonstrate the powerful generality of various code-related tasks, such as code generation \cite{chen2021evaluating,dong2023self,AceCoder,SCoT,yang2024improving,liu2024exploring}, program repair \cite{xia2023automated,10172854}, and code summarization \cite{10.1145/3551349.3559555,geng2023empirical}. However, as an alternative solution in automated code translation, their potential has not been exhaustively investigated yet.

\textbf{Empirical Study.} This work performs an empirical study on five recent LLMs, including GPT-3.5 \cite{ModelsOp75:online}, LLaMA-33B/13B/7B \cite{touvron2023llama}, and CodeGen \cite{nijkamp2022codegen}, for automated code translation task and compare their performance with the three state-of-the-art learning-based transpilers, including TransCoder \cite{roziere2020unsupervised}, TransCoder-IR \cite{szafraniec2022code}, and TransCoder-ST \cite{roziere2021leveraging}. We first manually clean the widely-used code translation dataset released by Roziere et al. \cite{roziere2020unsupervised} due to many errors and inconsistencies among its parallel corpus. Based on the cleaned dataset, we evaluate the above models under the metrics of Computational Accuracy (CA) and Exact Match Accuracy (EM Acc), where four settings of translation datasets (i.e., C++ to Java, Java to C++, C++ to Python, and Python to C++) are involved in this empirical study. Results demonstrate certain LLMs have outperformed state-of-the-art learning-based transpilers, and LLMs with more parameters tend to carry more powerful translation capabilities, showing that automated code translation with LLMs is promising. Nevertheless, LLMs still suffer some accuracy issues. To delve deep into these issues and find rescues to improve their performance further, taking GPT-3.5 as an example, we manually analyze 174 failures it made and partition them into various categories, e.g., failures concerning Syntax, Logic, API, etc., where 38.51\% of failures are induced by a lack of comprehension of the source programs, 14.94\% of them are brought by the missing instructions of explicit I/O types, and 41.38\% of them are caused by the ignorance of the discrepancies between source and target programs.     

\textbf{UniTrans.} Enlightened by the above findings, we propose a \textbf{Uni}fied code \textbf{Trans}lation framework, namely \textbf{UniTrans}, to unleash various LLMs' capabilities in this field. In general, \textbf{UniTrans} exploits auto-generated test cases\footnote{In this work, we generate test cases via \textbf{UniTrans}, which should be discriminated from the evaluation-purpose unit test/test suite provided by the dataset.} as extra information for LLMs to alleviate the aforementioned drawbacks. On the one hand, test cases imply the requirements of programs for code comprehension. Besides, I/O types can be easily labeled on test cases to complement the translation objective. In addition, test cases can also be executed to double-check or offer hints (e.g., error messages) to repair the translated programs, thereby alleviating failures brought about by neglecting discrepancies. Specifically, \textbf{UniTrans} consists of three phases, i.e., (1) the Test Case Generation Phase, (2) the Translation Augmentation Phase, and (3) the Translation Repair Phase. The Test Case Generation Phase leverages LLMs to generate a series of inputs with the \textit{Input Generation Prompt} for source programs. Afterward, valid inputs can be selected, and their corresponding outputs can also be obtained via the execution of source programs, thereby gathering test cases. Since test cases are composed of simple I/O pairs, practitioners can easily decorate I/O types on their corresponding positions and convert them to fit target programs via heuristics. As such, the Translation Augmentation Phase can exploit these test cases to improve code translation quality with the \textit{Translation Augmentation Prompt}. Following that, the translated programs are double-checked by the above test cases via execution, and unpassed ones are shipped to the Translation Repair Phase to extract error information and further repair with the \textit{Repair Prompt}. \textbf{UniTrans} also provides an option of iterative repair for users to fix bugs in multiple rounds based on feedback from test cases.

To evaluate the effectiveness and universal applicability of \textbf{UniTrans}, we experiment with three LLMs, namely GPT-3.5, LLaMA-13B, and LLaMA-7B, on all six settings of translation datasets between Python, Java, and C++, a total of 568 samples for each PL, where the translation between Python and Java is newly introduced to verify the generality of our findings among unseen translation datasets. Extensive experimental results demonstrate \textbf{UniTrans} substantially boosts the code translation efficacy of three tested LLMs on almost all translation datasets. To be specific, GPT-3.5 obtains average improvements of 4.02\% in terms of CA and 13.28\% in terms of EM Acc. LLaMA-13B achieves average improvements of 19.20\% and 36.42\% in terms of CA and EM Acc, respectively. LLaMA-7B demonstrates average improvements of 28.58\% and 71.22\% in terms of each metric in order. Furthermore, we carry out a series of ablation studies and discussion experiments to investigate the contribution and influence of \textbf{UniTrans}'s each module with various LLMs, showing that test cases throughout the whole life-cycle of \textbf{UniTrans} are critically important. The contributions of this paper can be summarized below:
\begin{itemize}

\item We manually and rigorously cleaned the widely-used code translation dataset, including parallel PLs of Python, Java, and C++, and made an explicit breakdown to record our cleaning process. The cleaned dataset and the breakdown table have been published in \cite{yz10191153:online}.

\item We carry out an empirical study to investigate the performance of recent LLMs on automated code translation and carefully analyze their prospects and limitations. Meanwhile, a series of invaluable findings are summarized.

\item We propose and evaluate \textbf{UniTrans}, a \textbf{Uni}fied code \textbf{Trans}lation framework applicable to various LLMs to unleash their power in code translation. Motivated by our findings, \textbf{UniTrans} introduces auto-generated test cases throughout the whole code translation framework via translation augmentation and repair. Comprehensive evaluations are conducted to examine the effectiveness of \textbf{UniTrans} both quantitatively and qualitatively, including ablation studies, discussion experiments, and case studies. 
We open-source \textbf{UniTrans} in \cite{yz10191153:online}. 
\end{itemize}
\vspace{-0.6em}
\section{Background and Related Work}
\subsection{Automated Code Translation}
Automated code translation tools aim to construct a function that approximates $f$ such that given the source program $x$, the target program $y=f(x)$. Early studies proposed various rule-based approaches, such as C2Rust \cite{immunant65:online} and CxGo \cite{gotransp89:online} to carry out translations from C to Rust and Go with manually crafted rules. However, they are language-dependent and extremely labor-intensive, as developers have to implement an enormous amount of translation rules for every function, object, and standard library of every language pair. Besides, the translated programs also suffer low readability and correctness \cite{nguyen2013lexical, 10.1109/ICSE48619.2023.00072, nguyen2015divide}. To this end, learning-based transpilers have been successively proposed during the last several years \cite{chen2018tree,karaivanov2014phrase,nguyen2016mapping,oda2015learning}. Owing to the unavailability of bilingual samples, state-of-the-art learning-based approaches \cite{roziere2020unsupervised,roziere2021leveraging,szafraniec2022code} leverage unsupervised/weakly supervised techniques to train their model with massive monolingual data. Although they achieved remarkable improvement against previous rule-based approaches, their limitations are still evident, i.e., (1) their performance is still insufficient for practical deployment, and (2) their training expenditures are very resource-consuming and expensive.
To combat the above limitations, LLMs are considered in this work due to their powerful generality on a broad spectrum of code-related tasks without the necessity of re-training/fine-tuning \cite{zhao2023survey,nashid2023retrieval,zhang2023repocoder}. Consequently, we explore the ability of various LLMs in the code translation task, along with summarizing their strengths and weaknesses. Moreover, we propose \textbf{UniTrans} to further unleash LLMs' code translation performance.

\subsection{Automated Test Case Generation}
Numerous literature has put forward many approaches to automatically generate test cases for a focal method during the last several decades. Traditional tools, such as Randoop \cite{pacheco2007feedback}, EvoSuite \cite{fraser2011evosuite}, and MOSA \cite{panichella2015reformulating}, exploit search-based heuristics to generate test cases, suffering limitations on diversity and quantity. A series of follow-up studies \cite{tufano2020unit,10.1145/3510003.3510141} proposed various learning-based approaches to overcome the above limitations, but they require massive training resources. Recent research leverages the generality of LLMs to generate test cases via zero-shot/few-shot learning, which has drawn lots of attention in academia and industry owing to their impressive performance and lightweight characteristic \cite{chen2022codet,lemieux2023codamosa,yuan2023no,tang2023chatgpt}. Nonetheless, owing to the nature of code translation, source programs are known. Therefore, we leverage source programs to validate the generated program inputs and gather corresponding outputs, thereby creating exactly correct test cases.

\subsection{Automated Program Repair} 
Automated Program Repair (APR) tools are designed to patch buggy code given the original code and corresponding buggy location. Traditional APR tools \cite{bader2019getafix, le2012systematic,martinez2016astor,ghanbari2019practical} are based on human-crafted heuristics or templates, leading to a lack of generality on unseen types of bugs. Subsequently, learning-based APR tools, such as Recoder \cite{zhu2021syntax}, DeepFix \cite{gupta2017deepfix}, and CODIT \cite{9181462}, emerged to generate more diverse and expressive patches. Nevertheless, owing to their enormous reliance on historical bug-fixing data, researchers nowadays tend to explore more lightweight approaches using LLMs without re-training/fine-tuning \cite{xia2023automated, xia2022less,xia2023conversational}. Different from the above approaches, given the auto-generated test cases for evaluation, our idea can employ the specific error information fetched from the program execution results to assist the program repair during the Translation Repair Phase.

\section{Motivation}
\label{Motivation}
This section introduces the motivation of \textbf{Unitrans}, which is derived from an in-depth analysis of failed cases of LLMs. In fact, we first conducted an empirical study on four translation datasets, i.e., C++ to Java, Java to C++, C++ to Python, and Python to C++, with two evaluation metrics, i.e., Computational Accuracy (CA) and Exact Match Accuracy (EM Acc) mentioned in Section \ref{Evaluation Metrics}, to explore the performance of various recent LLMs against state-of-the-art learning-based transpilers. The detailed experimental setting, dataset, and results are presented in Section \ref{Experimental Setting} and \ref{RQ1: What is the performance of recent LLMs against state-of-the-art learning-based transpilers in code translation?}. During the empirical study, we take the best-performing LLM (i.e., GPT-3.5) as an example and carry out the in-depth analysis based on a series of failed cases sampled from GPT-3.5's results, thereby exploring potential improvement directions for LLMs in code translation.

Specifically, to ensure a 95\% confidence level and 5\% confidence interval when sampling, we follow previous work \cite{chen2020comprehensive,zhang2019empirical, kotrlik2001organizational,xue2024automated} to randomly sample a total of 195 failed cases\footnote{Here, we define the failed cases as those that cannot pass all given unit tests, i.e., using CA as the assessment criterion.} from GPT-3.5's empirical results, where 52 are from C++ to Python, 43 are from Python to C++, 30 are from Java to C++, and 70 are from C++ to Java. Particularly, 21 cases extracted from the dataset of C++ to Java are ignored, as their mistakes are owing to the misintroduction of ``import'' statements or wrapped by ``class'', which can be easily eliminated by regular expressions. Eventually, 174 failed cases remained. Subsequently, we made a systematic taxonomy according to the cause of each failure. In detail, independent labeling is conducted by the first and second authors, followed by a double-check through a review process, and then a final taxonomy decision is reached through in-depth discussions. Following the above rigorous procedure, the remaining 174 failed cases are categorized into six classes. 

\begin{figure}[htbp]
\vspace{-0.3cm}
\setlength{\abovecaptionskip}{0.1cm}
\includegraphics[width=0.65\textwidth]
{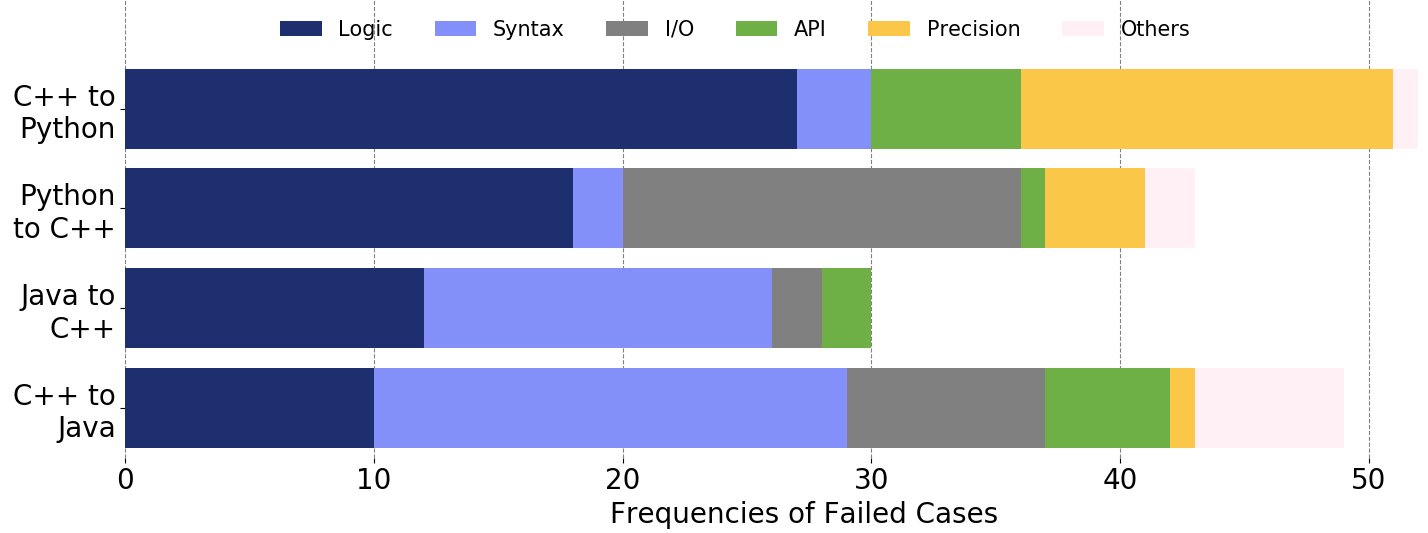}
\caption{Frequencies of Failed Cases in Each Translation Dataset}
\label{Frequencies of Failed Cases in Each Translation Dataset}
\vspace{-0.3cm}
\end{figure}

The detailed taxonomy, examples, and corresponding numbers are shown in Table \ref{Failure Cases Taxonomy}. Moreover, we present Figure \ref{Frequencies of Failed Cases in Each Translation Dataset} to showcase the frequencies of various failures in each translation dataset. Apparently, \textbf{Logic failure} is the most prominent category, accounting for 20.41\% to 51.92\% of all failures across diverse translation datasets, indicating that lack of comprehension of the source program is a primary weakness of GPT-3.5. The \textbf{I/O failure} occurs most frequently (37.21\%) when translating from Python to C++. Because dynamically typed PLs (e.g., Python) without explicit type declaration can hardly infer the parameters/return values' type of statically typed PLs (e.g., C++), showing that prompting I/O types is necessary when code translation. On the contrary, when translating from statically typed PLs to dynamically typed PLs, \textbf{mistakes concerning Precision} become apparent (28.85\%) owing to their differences in data types (e.g., C++ has double/float types, while Python only has the float type.) and operations (e.g., integer division as shown in Table \ref{Failure Cases Taxonomy}). Additionally, the \textbf{Syntax failure} mainly appears when translating between Java and C++, as these two PLs' syntaxes are more complicated compared with Python, leading to GPT-3.5 ignoring some discrepancies between source and target PLs. Similarly, \textbf{mistakes of API}, such as API mis-migration or misuse, are still typically due to the unfamiliarity of both PLs when translation. In summary, we attribute failures concerning Logic to the lack of comprehension of the source program (38.51\%). Besides, we argue the I/O failure is induced by missing explicit I/O type declaration (14.94\%). Finally, we credit mistakes relevant to Syntax, API, and Precision to the ignorance of the discrepancies between source and target PLs (41.38\%). 

Enlightened by the above findings in the manual analysis, we carry an idea of introducing test cases as extra information to alleviate the limitations of LLMs during their code translation, and the reasons are three-fold. Firstly, test cases imply the requirements of programs, facilitating LLMs' comprehension of the program logic. Besides, I/O types can be easily decorated on test cases, thereby mitigating the mistakes induced by missing I/O type instructions. Thirdly, executing test cases can double-check or provide hints (e.g., error messages) to rectify incorrectly translated programs, thereby alleviating mistakes brought by neglecting discrepancies between PLs. Consequently, \textbf{UniTrans} is proposed in this work, and its detailed introduction is elaborated in the following sections.

\vspace{-0.2cm}
\section{UniTrans}
\textbf{UniTrans} consists of three phases, namely (1) the Test Case Generation Phase, which leverages LLMs and source programs to generate test cases, (2) the Translation Augmentation Phase, employing auto-generated test cases as extra information to augment translation and inspect their correctness, and (3) the Translation Repair Phase, which further repairs incorrectly translated programs assisted by test case execution results. The overview of \textbf{UniTrans} is shown in Figure \ref{UniTrans}. We elaborate on each of its components as follows. 

\begin{table}[htbp]
\setlength{\abovecaptionskip}{0.1cm}
\caption{Failure Cases Taxonomy}
\label{Failure Cases Taxonomy}
\setlength\tabcolsep{1pt}
\scriptsize
\begin{threeparttable}
\begin{tabular}{lllll}
\toprule
\textbf{Category}  & \multicolumn{3}{c}{\textbf{Description and Examples}}       & \multicolumn{1}{c}{\textbf{Amount}} \\ \midrule
\multirow{6}{*}{Logic}     & \multicolumn{3}{l}{The translated program bears logical inconsistencies against the ground truth program.} & \multicolumn{1}{c}{\multirow{6}{*}{67}}     \\ \cmidrule{2-4}
          & \multicolumn{1}{c}{\textbf{Source Program}}  & \multicolumn{1}{c}{\textbf{Translated Program}}  & \multicolumn{1}{c}{\textbf{Ground Truth Program}}  &                            \\ \cmidrule(lr){2-2} \cmidrule(lr){3-3}
 \cmidrule(lr){4-4}  
          & \begin{tabular}[c]{@{}l@{}}return * max\_element(arr, arr+n); \\ //source C++ code \\ //gets the max value of the top-n items.\end{tabular} & \begin{tabular}[c]{@{}l@{}}return Arrays.stream(arr).max().getAsInt(); \\ //incorrect Java code\\ \textcolor[rgb]{1,0,0}{//gets the max value of the whole array.} \\ \end{tabular} & \begin{tabular}[c]{@{}l@{}}Arrays.sort(arr, 0, n);\\ return arr{[}n-1{]};\\ //ground truth Java code\end{tabular} &                            \\
          \midrule
\multirow{6}{*}{Syntax}    & \multicolumn{3}{l}{Incorrectly introduces the syntax of the source PL to the target PL or makes the syntactic errors in the target PL.}                & \multicolumn{1}{c}{\multirow{6}{*}{38}}     \\ \cmidrule{2-4}
          & \multicolumn{1}{c}{\textbf{Source Program}} & \multicolumn{1}{c}{\textbf{Translated Program}}  & \multicolumn{1}{c}{\textbf{Ground Truth Program}}   &                            \\ \cmidrule(lr){2-2} \cmidrule(lr){3-3}
 \cmidrule(lr){4-4}          & \begin{tabular}[c]{@{}l@{}}return ( ! ( x / z ) ) ? x : z;\\ //source C++ code\end{tabular}    & \begin{tabular}[c]{@{}l@{}}return ( ! ( x / z ) ) ? x : z;\\ //incorrect Java code\\ \textcolor[rgb]{1,0,0}{//``!'' is not applicable for digits in Java.}\end{tabular}            & \begin{tabular}[c]{@{}l@{}}return ( ( x / z ) == 0 ) ? x : z ;\\ //ground truth Java code\end{tabular}            &                            \\ 
 \midrule
\multirow{6}{*}{I/O}       & \multicolumn{3}{l}{The translated program carries inconsistent I/O types against the ground truth program.}     & \multicolumn{1}{c}{\multirow{6}{*}{26}}     \\ \cmidrule{2-4}
          & \multicolumn{1}{c}{\textbf{Source Program}} & \multicolumn{1}{c}{\textbf{Translated Program}}  & \multicolumn{1}{c}{\textbf{Ground Truth Program}}  &                            \\ \cmidrule(lr){2-2} \cmidrule(lr){3-3} \cmidrule(lr){4-4}
& \begin{tabular}[c]{@{}l@{}}def evenlength(n):\\ \#source Python code\end{tabular}                                                                    & \begin{tabular}[c]{@{}l@{}}int evenlength(int n)\{\\ //incorrect C++ code\\ \textcolor[rgb]{1,0,0}{//incorrect input/output types.}\end{tabular}                                   & \begin{tabular}[c]{@{}l@{}}string evenlength ( string n ) \{\\ //ground truth C++ code\end{tabular}               &                            \\ 
\midrule
\multirow{7}{*}{API}       & \multicolumn{3}{l}{Incorrectly duplicates the API symbols of the source PL to the target PL or misuses APIs of the target PL.}                                      & \multicolumn{1}{c}{\multirow{7}{*}{14}}     \\ \cmidrule{2-4}
          & \multicolumn{1}{c}{\textbf{Source Program}} & \multicolumn{1}{c}{\textbf{Translated Program}}  & \multicolumn{1}{c}{\textbf{Ground Truth Program}}   &    \\ \cmidrule(lr){2-2} \cmidrule(lr){3-3} \cmidrule(lr){4-4}
& \begin{tabular}[c]{@{}l@{}}stack <int> s;\\...\\int tp = s.top();\\//source C++ code\end{tabular}                                                         & \begin{tabular}[c]{@{}l@{}}Stack<Integer> s = new Stack<>();\\...\\int tp = s.top();//incorrect Java code\\ \textcolor[rgb]{1,0,0}{//stack::top() is only available in C++}\end{tabular}              & \begin{tabular}[c]{@{}l@{}}Stack<Integer> s = new Stack<>();\\...\\int tp = s.peek();\\//ground truth Java code\end{tabular}           &                            \\ 
\midrule
\multirow{7}{*}{Precision} & \multicolumn{3}{l}{The translated program returns digits of inconsistent precision against the ground truth program.}  & \multicolumn{1}{c}{\multirow{7}{*}{20}}     \\ \cmidrule{2-4}& \multicolumn{1}{c}{\textbf{Source Program}} & \multicolumn{1}{c}{\textbf{Translated Program}}  & \multicolumn{1}{c}{\textbf{Ground Truth Program}} & \\ \cmidrule(lr){2-2} \cmidrule(lr){3-3} \cmidrule(lr){4-4}
          & \begin{tabular}[c]{@{}l@{}}return (b/m-1)*(b/m)/2;\\ \#source C++ code\end{tabular} & \begin{tabular}[c]{@{}l@{}}return (b/m-1)*(b/m)/2\\ \#incorrect Python code\\ \textcolor[rgb]{1,0,0}{\#as `b' is an integer parameter, integer} \\ \textcolor[rgb]{1,0,0}{\#division (`//') should be used instead.}\end{tabular} & \begin{tabular}[c]{@{}l@{}}return (b//m-1)*(b//m)//2\\ \#ground truth Python code\end{tabular}      &                            \\ 
\midrule
Others    & \multicolumn{3}{l}{Other mistakes.}  & \multicolumn{1}{c}{9}   \\
\midrule
Total & & & & \multicolumn{1}{c}{174} \\
\bottomrule
\end{tabular}
\begin{tablenotes}
\definecolor{light cyan}{HTML}{E1FFFF}
\fboxsep1.5pt
\item[$\Phi$] \scriptsize Only code fragments are listed in examples. Comments highlighted in \textcolor[rgb]{1,0,0}{red} are the specific errors of the translated programs.
\end{tablenotes}
\end{threeparttable}
\vspace{-1.3em}
\end{table}

\begin{figure}[htbp]
\setlength{\abovecaptionskip}{0.1cm}
\includegraphics[width=1\textwidth]
{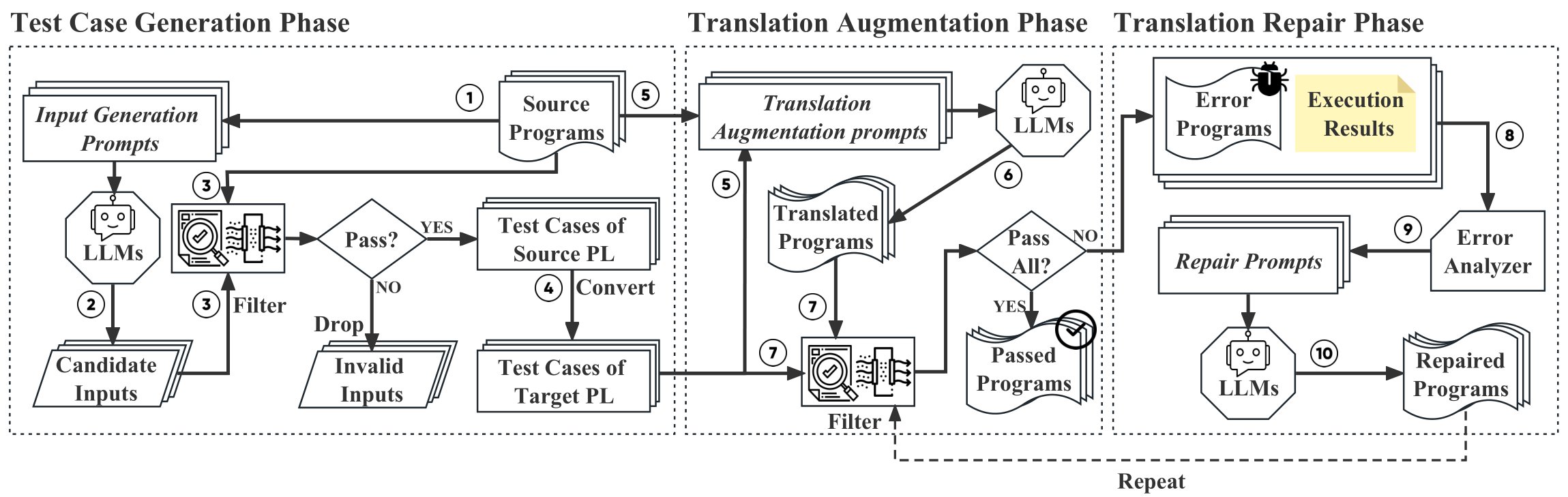}
\caption{UniTrans}
\label{UniTrans}
\vspace{-1em}
\end{figure}

\subsection{Test Case Generation}
\label{Test Case Generation}
\begin{figure}[htbp]
\vspace{-0.2cm}
\setlength{\abovecaptionskip}{0.1cm}
\includegraphics[width=0.7\textwidth]
{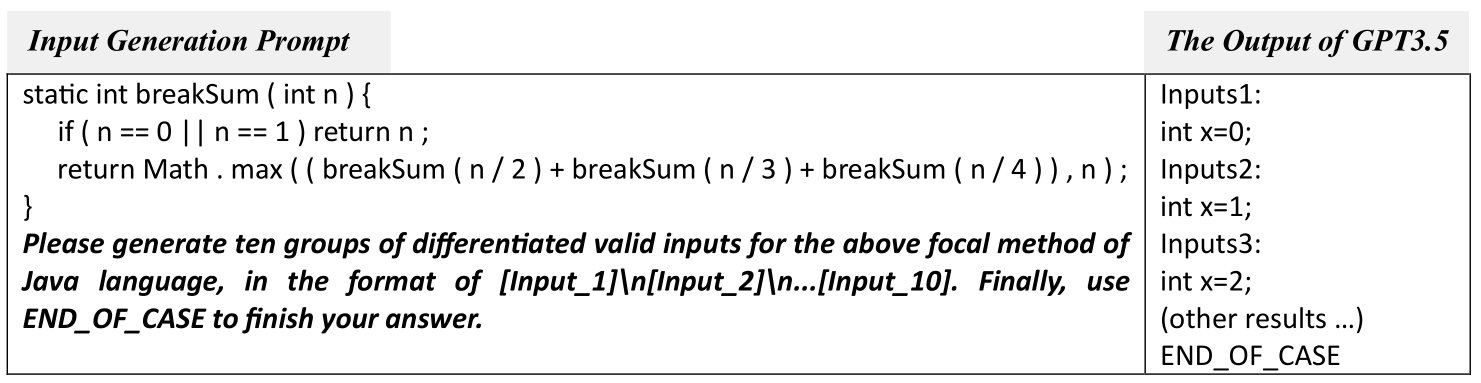}
\caption{An Example of Input Generation Prompt}
\label{Input Generation Prompt}
\vspace{-0.5cm}
\end{figure}

\begin{figure}[htbp]
\setlength{\abovecaptionskip}{0.1cm}
\includegraphics[width=0.6\textwidth]
{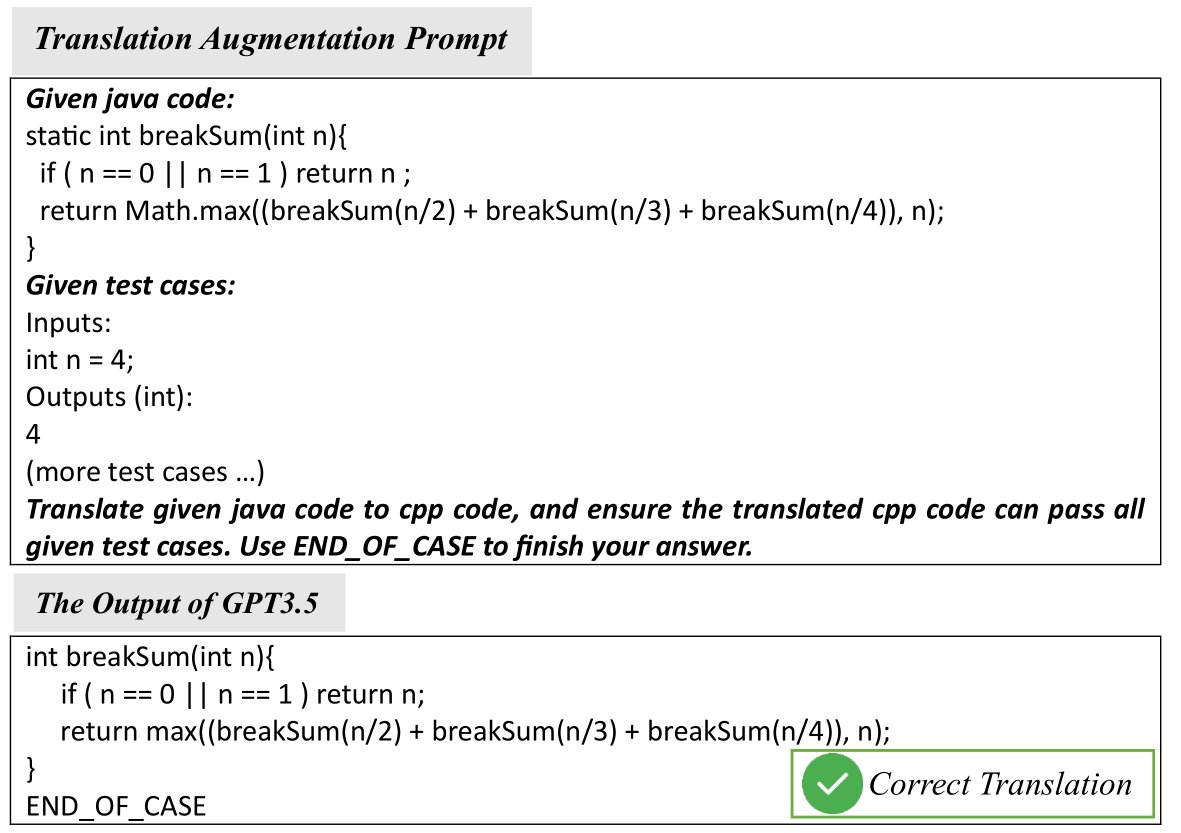}
\caption{An Example of Translation Augmentation Prompt}
\label{An Example of Translation Augmentation Prompt}
\vspace{-0.5cm}
\end{figure}

Test Case Generation Phase is presented on the left-hand side of Figure \ref{UniTrans}.
Instead of directly generating test cases for focal methods, we leverage LLMs to generate methods' candidate inputs first (Step 1-2) with the \textit{Input Generation Prompt}. This prompt can be formally defined as ``\$\{$prog_{src}$\}\textbackslash nPlease generate ten groups of differentiated valid inputs for the above focal method of \$\{$pl_{src}$\} language, in the format of [Input\_1]\textbackslash\textbackslash n[Input\_2]\textbackslash\textbackslash n...[Input\_10]. Finally, use END\_OF\_CASE to finish your answer.'', where \$\{$prog_{src}$\} and \$\{$pl_{src}$\} are placeholders for a source program and the name of source PL (e.g., Java and Python). An example is shown in Figure \ref{Input Generation Prompt}, where the text in bold and italic format are natural language instructions 
, the same as the follow-up prompts shown in Figures \ref{An Example of Translation Augmentation Prompt}, \ref{An Example of Repair Prompt}, and \ref{Basic Prompt}. Particularly, we instruct LLMs to generate 10 candidate inputs for each inference, explicitly stating the requirement that inputs need to be differentiated.
Afterward, we use source programs as calibrations to filter valid inputs and obtain their corresponding outputs via execution (Step 3), thereby constructing collections of test cases $tc_{src}$. Before pipelining to the second phase, we convert their formats to fit the target PLs via heuristic rules, thereby formulating $tc_{tar}$ (Step 4). $tc_{tar}$ can be arranged as ``Inputs:\textbackslash n\$\{$inp_{tar}$\}Outputs:\textbackslash n\$\{$out_{tar}$\}'', where \$\{$inp_{tar}$\} and \$\{$out_{tar}$\} are the placeholders of inputs and outputs of a test case in the format of the target PL. In particular, for target PLs with statically typed characteristics, such as Java and C++, we complement $tc_{tar}$ with input/output types, i.e., ``Inputs:\textbackslash n\$\{$inp_{tar}^{*}$\}Outputs (\$\{$type_{tar}$\}):\textbackslash n\$\{$out_{tar}$\}'', where \$\{$inp_{tar}^{*}$\} is the placeholder of a test case's inputs with explicit variable types, while \$\{$type_{tar}$\} is the placeholder of outputs' type of a test case. Both of them are in the target PL's format.

\subsection{Translation Augmentation}
After the Test Case Generation Phase, \textit{Translation Augmentation Prompt}s can be composed with prepared test cases and source programs (Step 5), as an example shown in Figure \ref{An Example of Translation Augmentation Prompt}. The prompt template can be formally defined as ``Given \$\{$pl_{src}$\} code:\textbackslash n\$\{$prog_{src}$\}\textbackslash nGiven test cases:\$\{$TC_{tar}$\}\textbackslash nTranslate given \$\{$pl_{src}$\} code to \$\{$pl_{tar}$\} code, and ensure the translated \$\{$pl_{tar}$\} code can pass all given test cases. Use END\_OF\_CASE to finish your answer.'', where \$\{$TC_{tar}$\} consists of a series of prepared test cases $tc_{tar}$, \$\{$pl_{tar}$\} is the placeholder for the name of the target PL.
Hence, LLMs can leverage the above prompt to augment the code translation with extra test case information (Step 6). 
Afterward, we further execute the translated programs with prepared test cases to double-check their correctness as a preliminary inspection (Step 7). A program that passes all given test cases is deemed correct, and we return it to users. Otherwise, we will pipeline this sample to the last phase for repair. The Translation Augmentation Phase is listed in the middle of Figure \ref{UniTrans}. 

\subsection{Translation Repair}
\begin{figure}[htbp]
\vspace{-0.3cm}
\setlength{\abovecaptionskip}{0.1cm}
\includegraphics[width=0.8\textwidth]
{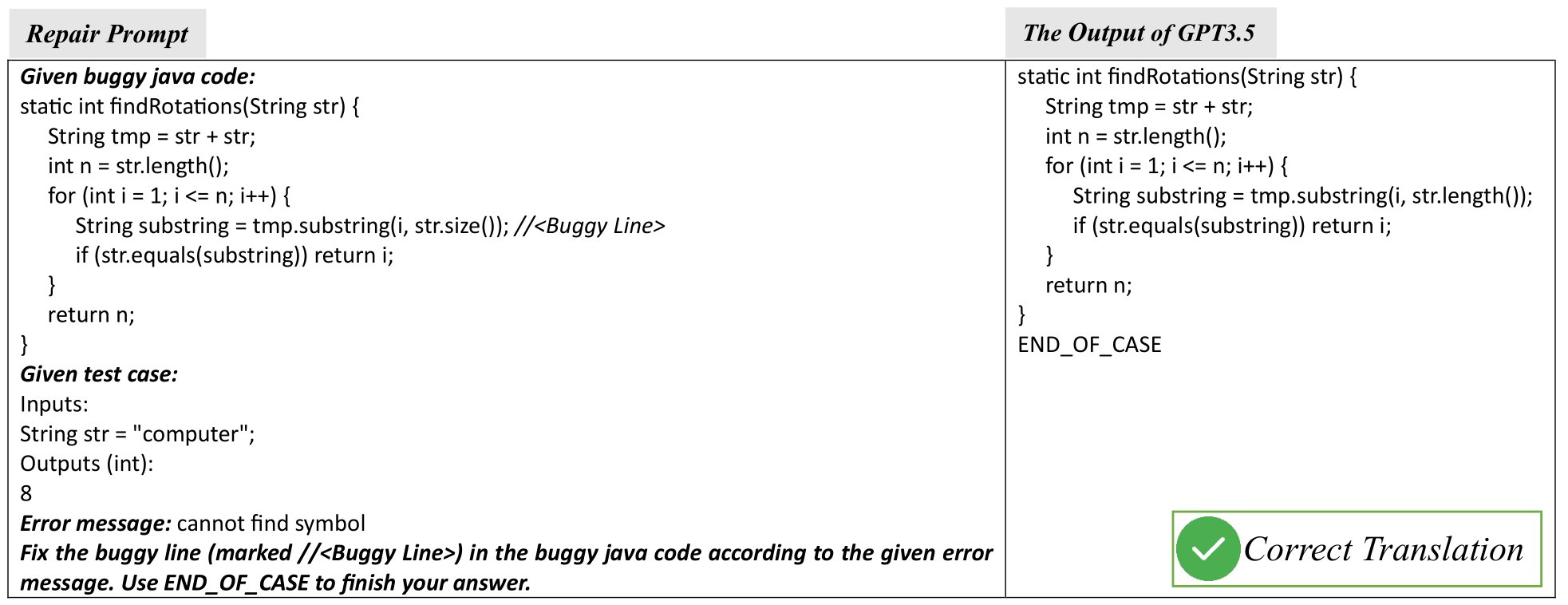}
\caption{An Example of Repair Prompt}
\label{An Example of Repair Prompt}
\vspace{-0.3cm}
\end{figure}

\begin{figure}[htbp]
\vspace{-0.3cm}
\setlength{\abovecaptionskip}{0.1cm}
\includegraphics[width=0.6\textwidth]
{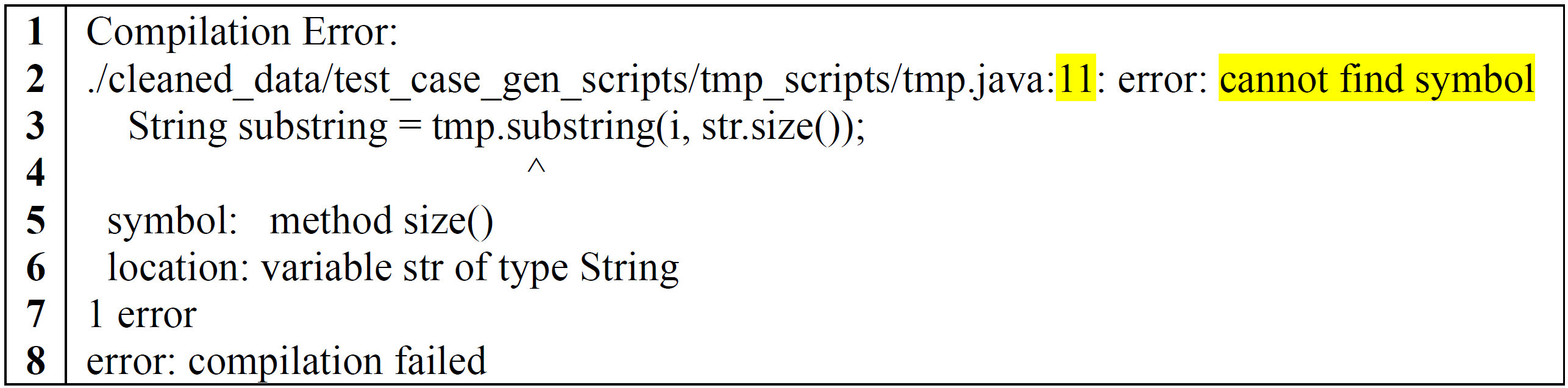}
\caption{An Execution Result Example}
\label{An Execution Result Example}
\vspace{-0.5cm}
\end{figure}

Translated programs once failed on certain test cases, their compilation/runtime errors will be thrown via the stack backtraces, while logic errors will list the explicit discrepancies between the expected outputs and actual outputs to identify the output inconsistencies. To this end, as shown in the right-hand side of Figure \ref{UniTrans}, we design an Error Analyzer (EA) to extract error information from the above Execution Results (Step 8), including error lines and error messages. Taking the buggy Java program in Figure \ref{An Example of Repair Prompt} as an example, Figure \ref{An Execution Result Example} is its Execution Result throwing a compilation error. Following our designed regular expressions, EA can extract the error line number (``11'') and error message (``cannot find symbol'') in Line 2, thereby providing necessary information in the TRP for program repair (highlighted in yellow). It should be noted that, as each translated program needs to be inserted into the testing template file (e.g., tmp.java and tmp.py, refer to \cite{yz10191153:online} for details) for evaluation, ``11'' here is the buggy line number in tmp.java rather than in the translated program. Thus, we must subtract 6 prefix lines of our translated program (i.e., 11-6=5) in tmp.java. As such, we can precisely specify the buggy line in the translated program, as shown in Figure \ref{An Example of Repair Prompt}. Besides, Execution Results of different PLs and errors (e.g., compilation, runtime, and logic errors) are slightly different. To that end, we tailor different regular expressions towards various PLs and error types in EA and have published them in \cite{yz10191153:online}.

Afterward, the extracted information can be embedded into the \textit{Repair Prompt}s, facilitating LLMs to fix corresponding error programs and refine the translation (Step 9-10). The prompt for compilation/runtime error can be formally defined as ``Given buggy \$\{$pl_{tar}$\} code:\textbackslash n\$\{$prog_{tar}^{*}$\}\textbackslash n Given test case:\$\{$tc_{tar}^{*}$\}\textbackslash nError message: \$\{$err\_msg$\}\textbackslash nFix the buggy line (marked \$\{$com\_sym_{tar}$\} <Buggy Line>) in the buggy \$\{$pl_{tar}$\} code according to the given error message. Use END\_OF\_CASE to finish your answer.'', where \$\{$prog_{tar}^{*}$\}, \$\{$tc_{tar}^{*}$\}, \$\{$err\_msg$\}, and \$\{$com\_sym_{tar}$\} are the placeholders of a translated program with a buggy line specified, a failed $tc_{tar}$, an error message, and the comment symbol of the target PL (e.g., ``//'' for Java and C++ while ``\#'' for Python) in order. It should be noted that even though each translated program can be evaluated with more than one test case, we only feed the first failed case for repair due to the restriction of the prompt token length. The definition of prompt template of logic error is slightly different, which is: ``Given buggy \$\{$pl_{tar}$\} code:\textbackslash n\$\{$prog_{tar}$\}\textbackslash nGiven test case:\$\{$tc_{tar}^{*}$\}\textbackslash nError message: \$\{$err\_msg$\}\textbackslash nFix the buggy \$\{$pl_{tar}$\} code according to the given error message. Use END\_OF\_CASE to finish your answer.'', where \$\{$prog_{tar}$\} is the placeholder of an incorrectly translated program without a specified buggy line, as logic errors are hard to locate. 
Besides, it is also optional for users to repeatedly evaluate the repaired programs with auto-generated test cases to fetch updated error information and repair until reaching the pre-defined maximum iteration, namely iterative repair. 

\section{Experimental Setting}
\label{Experimental Setting}
This section elaborates on the whole experimental setting across this work, including the dataset, models, metrics, implementation, research questions, and evaluation methodology.
\subsection{Data Cleaning} 
\label{Data Cleaning}
To assess the effectiveness of TransCoder, Roziere et al. \cite{roziere2020unsupervised} released a widely-used translation dataset composed of 948 parallel functions in Python, Java, and C++. However, only 568 of them contain evaluation-purpose unit tests for at least one PL, where 464 unit tests are for Python, 482 are for Java, and 467 are for C++. As such, we focus on this part of the dataset in our experiments. 
The dataset was collected from GeeksforGeeks \cite{Geeksfor48:online}, which is an online platform containing many coding problems and solutions in various PLs. However, based on our observation of the preliminary experiments on GPT-3.5, lots of failures are induced by data noises in unit test scripts or inconsistencies among parallel functions. 
To reveal the actual ability of different approaches, four authors with more than three years of programming experience in Python, Java, and C++ are involved in cleaning the dataset manually. All their corrected samples are required to be cross-checked by two participators, thereby assuring the data quality. 
For clarification, we divide those identified noises into different categories, such as logic inconsistency, runtime/compilation error, and unit test error. In summary, a total of 252 errors were found, where 132 errors in Python, 58 errors in Java, and 62 errors in C++. The detailed breakdown of these data noises is attached in our replication package due to the page limit. All these noises are eliminated after the data cleaning.

\subsection{Studied Models}
\label{Studied Models}
We introduce five recent LLMs of diverse sizes and families as well as three state-of-the-art learning-based transpilers for empirical study. Their detailed information is listed below.

$\bullet$ \textbf{GPT-3.5 \cite{ModelsOp75:online}:} A set of models improved on GPT-3, fine-tuned with RLHF techniques for understanding human instructions and generating natural language and code. We use OpenAI's APIs to access its latest version, i.e., gpt-3.5-turbo-0613. 

$\bullet$ \textbf{LLaMA \cite{touvron2023llama}:} A family of multilingual large language models ranging from 7B to 65B in parameter size, trained from various open-source datasets of up to 1.4 trillion tokens, where 4.5\% are code from Github. We include three versions of 7B/13B/33B for experiments.

$\bullet$ \textbf{CodeGen \cite{nijkamp2022codegen}:} A collection of language models trained on up to 354.7 billion English tokens and 222.5 billion code tokens. It also has diverse sizes in parameters ranging from 350M to 16.1B. During the experiment, we use codegen-6B-multi.

$\bullet$ \textbf{TransCoder \cite{roziere2020unsupervised}:} It is an unsupervised machine translation-based transpiler pre-trained with cross-lingual language modeling, denoising auto-encoding, and back-translation. As such, a vast amount of monolingual samples can be leveraged for training. TransCoder's translation ability covers Python, Java, and C++.

$\bullet$ \textbf{TransCoder-IR \cite{szafraniec2022code}:} It is an incremental work of TransCoder, introducing low-level compiler Intermediate Representation (IR) to improve the code translation performance. Along with the pre-training tasks of TransCoder, TransCoder-IR involves extra pre-training tasks, such as translation language modeling, translation auto-encoding, and IR generation. TransCoder-IR was trained for translation among Java, C++, Rust, and Go.

$\bullet$ \textbf{TransCoder-ST \cite{roziere2021leveraging}:} It is another augmented version based on TransCoder, which leverages automatically generated test cases to filter out invalid translations from the back-translation process, thereby improving the performance of unsupervised machine translation approaches in the code translation task. 

\subsection{Evaluation Metrics}
\label{Evaluation Metrics}
Following the previous studies in the code translation field \cite{roziere2020unsupervised, roziere2021leveraging,10.1109/ICSE48619.2023.00072,szafraniec2022code}, we adopt Computational Accuracy (CA) and Exact Match Accuracy (EM Acc) to assess the performance of each model.

\textbf{Exact Match Accuracy (EM Acc):} It computes the ratio of translations that exactly match ground truths, which can be formally defined as: 
\begin{equation}
    EM \ Acc = \frac{\sum^{N}_{k=1}em(y_{k},\hat{y_{k}})}{N}, \ \textbf{where} \ em(y_{k},\hat{y_{k}}) = \left\{\begin{matrix}
1 & y_{k}=\hat{y_{k}} \\
0 & y_{k}\neq\hat{y_{k}} \\
\end{matrix}\right. \label{eq1}
\end{equation}
where $N$ denotes the total number of translation samples, $y_k$ denotes the ground truth of the $k$-th sample, and $\hat{y_{k}}$ denotes the translated program via a certain transpiler for the $k$-th sample. For example, given $y_k$ and $\hat{y_{k}}$, only if they are exactly identical, they can be deemed a correct translation in EM Acc. Thus, EM Acc concludes the lower-bound of the effectiveness of transpilers.

\textbf{Computational Accuracy (CA):} It computes the ratio that the translated programs can produce the same execution result as the ground truths, given the same inputs. Thus, this metric considers the semantic equivalency of programs, which can be formally defined as:
\begin{equation}
CA = \frac{\sum^{N}_{k=1}ca(y_{k},\hat{y_{k}})}{N}, \ \textbf{where} \ ca(y_{k},\hat{y_{k}}) = \left\{\begin{matrix}
1 & Exec_{k}(y_k)=Exec_{k}(\hat{y_{k}})\\
0 & Exec_{k}(y_k)\neq Exec_{k}(\hat{y_{k}}) \\
\end{matrix}\right. \label{eq2}
\end{equation}
where $N$, $y_k$, and $\hat{y_{k}}$ carry the same meaning as those in EM Acc, $Exec_{k}(\cdot)$ denotes the execution result of a program with the test suite of the $k$-th sample. For example, even if $y_k$ and $\hat{y_{k}}$ are not identical literally, they are considered a correct translation in CA, as long as $Exec_{k}(y_k)=Exec_{k}(\hat{y_{k}})$.    
\subsection{Implementation}
\label{Implementations}
\begin{figure}[htbp]
\vspace{-1em}
\setlength{\abovecaptionskip}{0.1cm}
\includegraphics[width=0.6\textwidth]
{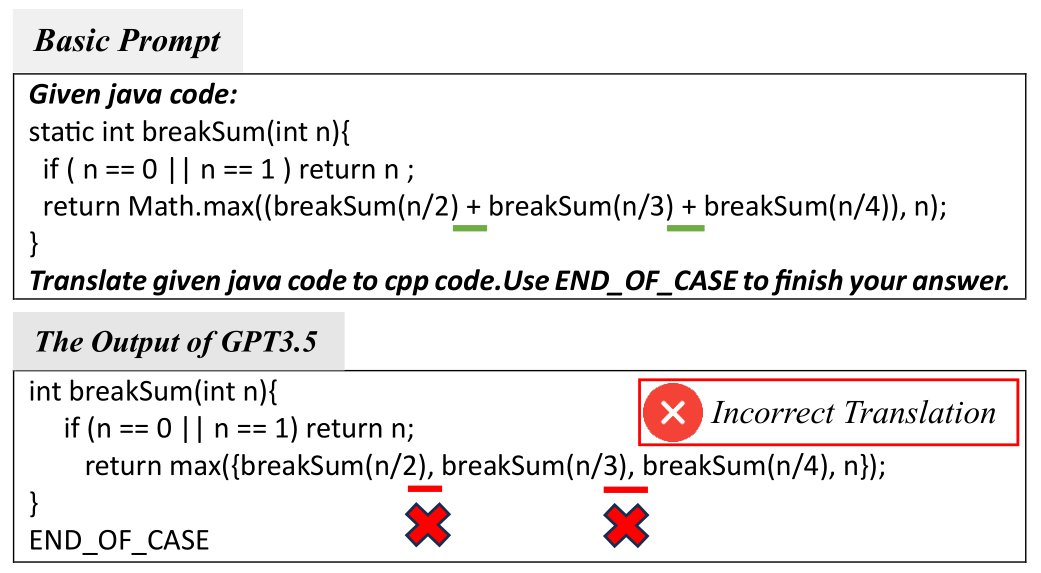}
\caption{Basic Prompt}
\label{Basic Prompt}
\vspace{-2em}
\end{figure}

Regarding LLMs, we implement GPT-3.5 by invoking OpenAI's API \cite{platform32:online}. For open-source LLMs, such as LLaMA and CodeGen, we instantiate them with their replication packages and load their weights from HuggingFace \cite{HuggingF70:online}. The default settings of LLMs are the same, using nucleus sampling \cite{holtzman2019curious} with $top\_p$=0.95, $temperature$=0.8, and 10 samples per translation. Besides, we design a \textit{Basic Prompt} to evaluate LLMs' code translation capability, which can be formally defined as ``Given \$\{$pl_{src}$\} code:\textbackslash n\$\{$prog_{src}$\}\textbackslash nTranslate given \$\{$pl_{src}$\} code to \$\{$pl_{tar}$\} code. Use END\_OF\_CASE to finish your answer.'' An example of the \textit{Basic Prompt} and a generated result is shown in Figure \ref{Basic Prompt}. In experiments, we adopt a one-shot learning setting to make every LLM understand more about the task and only evaluate their first outputs.
For learning-based translators, we reproduce them using their published source code and load them with the optimal weights for each translation dataset. Following that, we fix their $beam\_size$s=10 and fetch their first outputs for re-evaluation during inference on the cleaned dataset.

Towards \textbf{UniTrans}, the settings of its backbone LLMs are kept as default. In the Translation Augmentation Phase, considering the trade-off between performance and efficiency, we randomly select 3 generated test cases to construct \textit{Translation Augmentation Prompt}s. Besides, in the Translation Repair Phase, we leverage LLMs to repair error programs for 1 round as the default setting, considering the efficiency issue. 

\subsection{Research Questions and Evaluation Methodology}
\label{Research Questions and Evaluation Methodology}
Here, we introduce six research questions of this work and propose our evaluation methodology.

\textbf{RQ1: What is the performance of recent LLMs against state-of-the-art learning-based transpilers in code translation?} This RQ extensively assesses the code translation performance of recent LLMs. In this empirical study, we evaluate the models mentioned in Section \ref{Studied Models} on four translation datasets, i.e., C++ to Python, Python to C++, Java to C++, and C++ to Java, leaving the remaining two datasets, i.e., Python to Java and Java to Python, intact so as to examine the generality of \textbf{UniTrans} on translation scenarios that have not been analyzed in follow-up experiments. 
As such, we can quantitatively explore the practicality of LLMs in the code translation task and make a fair comparison with state-of-the-art learning-based transpilers.

\textbf{RQ2: How does UniTrans perform with different LLMs?} This RQ aims to examine the effectiveness of our proposal and its generalization capability on various LLMs. In experiments, different LLMs of diverse sizes are selected for investigation, including LLaMA-7B, LLaMA-13B, and GPT-3.5. 
For experimental datasets, along with the four pairs of translations, i.e., C++ to Python, Python to C++, Java to C++, and C++ to Java, used in Section \ref{RQ2: How does UniTrans perform with different LLMs?} (RQ1), we also newly include two translation datasets between Python and Java. Because the former four pairs of translation datasets have been analyzed in Sections \ref{Motivation} and \ref{RQ2: How does UniTrans perform with different LLMs?}, and we tailored specific methods to overcome LLMs' failure on them, thereby proposing UniTrans. If we only conduct experiments on these four datasets, we cannot know if our proposal has generality or not to other translation pairs.


\textbf{RQ3: What are the contributions of each component in UniTrans}? \textbf{UniTrans} consists of two components to boost its code translation performance, i.e., (1) the Translation Augmentation Phase (TAP) and (2) the Translation Repair Phase (TRP). 
Based on the experimental setting of RQ2, we further deconstruct \textbf{UniTrans} and successively add the TAP and TRP to LLMs \footnote{Before equipping with TAP and TRP, we have already employed the Test Case Generation Phase for LLMs to generate their respective test cases with the same setting as RQ2.}, including LLaMA-7B, LLaMA-13B and GPT-3.5, to investigate their contributions in terms of CA and EM Acc on each translation dataset.
    
\textbf{RQ4: How does the number of test cases influence the performance of UniTrans}? Incorporating more test cases in the \textit{Translation Augmentation Prompt} (i.e., in the TAP) indeed provides more references but occupies the valid prompt capacity and induces unexpected redundancy. Besides, this effect also carries certain impacts on the subsequent TRP, as potentially more vulnerabilities can be detected and repaired by adding more test cases. As such, to uncover an appropriate setting of this component, we investigate the performance of different LLMs with different numbers of test cases in this RQ. Due to the page limit, we consider Java as the target PL and divide four trials, where we randomly select $\{0, 1, 3, 5\}$ Java test cases for each group, respectively, from the generated test case pool of each experimented LLM (i.e., LLaMA-7B and GPT-3.5). We do not continue sampling more test cases because the line coverage increment has been gradually leveling off, as tested by a Java coverage tool, namely EMMA \cite{EMMAafre97:online}. Following that, we conduct experiments on the TAP and TRP, using diverse settings of test cases, and analyze the outcomes.

\textbf{RQ5: What is the performance of LLMs on the valid input generation task?} Efficiently generating valid inputs for programs facilitates the follow-up test case gathering, translation augmentation, and repair. Hence, this RQ aims to explore the performance of LLMs in generating valid inputs given limited attempts, thereby concluding the proper number of attempts for the guidance of practical deployment. Following the above experimental setting, we further discuss the LLMs' capabilities to generate at least $\{1, 3, 5\}$ valid inputs for each PL (i.e., Python, Java, and C++) given limited attempts of inference (i.e., $\{1,2,3\}$). As we mentioned in Section \ref{Test Case Generation}, for each inference, LLMs are instructed to generate 10 candidate inputs with the \textit{Input Generation Prompt}. Thus, they will totally generate $\{10, 20, 30\}$ candidate inputs in theory for each trial. Both LLaMA-7B and GPT-3.5 with the default setting are included in this experiment.

\textbf{RQ6: How do the rounds of iterative repair affect the performance of UniTrans}? 
Iteratively repairing might bring the buggy code closer to the ground truths but also may lead it to degenerate. Besides, endlessly repairing incurs extra unaffordable time costs. Hence, this RQ intends to unveil an appropriate maximum repair iteration of \textbf{UniTrans}. For this experiment, we pre-define four maximum repair iterations (i.e., $\{0,1,2,3\}$) and include two LLMs (i.e., LLaMA-7B and GPT-3.5). In detail, we instruct LLMs to consecutively repair 3 rounds and record their repair results sequentially on each translation dataset. Besides, we exclude EM Acc from the evaluation metrics because it underestimates the efficacy of the repair process and makes it hard to observe the performance differences, as discussed in Section \ref{RQ3: What are the contributions of each component in UniTrans?}. Meanwhile, CA also cannot capture all the repair details. For example, if a post-repair program can pass\footnote{Passing a unit test means the translated program can produce the same output as the ground truth program, given the same input.} more unit tests than the pre-repair program, the repair should be deemed effective even if it cannot pass the whole test suite. Whereas CA does not consider such improvement.
    
Thus, Apart from CA, we also introduce a new evaluation metric, namely Pass Rate (PR), to assess the performance variation of iterative repair in a more fine-grained manner. Specifically, PR measures the average percentage of unit tests that are passed by repaired programs. We consider PR as an auxiliary metric of CA in this experiment, as CA considers the whole test suite as an entity while PR can focus on individuals of the test suite. Therefore, when two counterparts achieve the same CA score, we use PR to distinguish their difference in the granularity of individual unit tests. 

\begin{table}[htbp]
\vspace{-0.5em}
\setlength{\abovecaptionskip}{0.1cm}
\caption{Empirical Results of Each Model for Code Translation}
\label{Empirical Results of Each Model for Code Translation}
\setlength\tabcolsep{3.5pt}
\scriptsize
\begin{threeparttable}
\begin{tabular}{lcccccccccc}
\toprule
\multirow{2.5}{*}{\textbf{Models}}               & \multicolumn{2}{c}{\textbf{C++ to Python}}   & \multicolumn{2}{c}{\textbf{Python to C++}}  & \multicolumn{2}{c}{\textbf{Java to C++}}     & \multicolumn{2}{c}{\textbf{C++ to Java}}  & \multicolumn{2}{c}{\textbf{Average}}    \\   \cmidrule(lr){2-3} \cmidrule(lr){4-5} \cmidrule(lr){6-7} \cmidrule(lr){8-9} \cmidrule{10-11}
\multicolumn{1}{c}{} & \textbf{CA}               & \textbf{EM Acc}           & \textbf{CA}               & \textbf{EM Acc}          & \textbf{CA}               & \textbf{EM Acc}           & \textbf{CA}               & \textbf{EM Acc} & \textbf{CA}               & \textbf{EM Acc}           \\ \midrule
\multicolumn{11}{c}{\textbf{Large Language Models}}\\ \midrule
CodeGen           & 30.17\%          & 4.40\%           & 22.70\%          & 1.94\%          & 35.12\%          & 7.57\%           & 11.20\%          & 2.46\% & 24.80\% & 4.09\%           \\
LLaMA-7B             & 23.49\%          & 3.17\%           & 16.49\%          & 1.41\%          & 32.98\%          & 10.04\%          & 21.16\%          & 9.86\% & 23.53\% & 6.12\%        \\
LLaMA-13B            & 33.19\%          & 4.05\%           & 32.98\%          & 2.46\%          & 37.90\%          & 10.39\%          & 40.25\%          & 12.15\% & 36.08\% & 7.26\%         \\
LLaMA-33B            & 47.63\%          & 4.75\%           & 38.12\%          & 1.94\%          & 59.53\%          & 17.43\%          & 45.23\%          & 12.50\% & 47.63\% & 9.16\%         \\
GPT-3.5          & \textbf{87.07\%} & \textbf{11.44\%} & \textbf{89.51\%} & \textbf{6.69\%} & \textbf{92.93\%} & \textbf{27.46\%} & \textbf{82.16\%} & \textbf{26.58\%} & \textbf{87.92\%} & \textbf{18.04\%}\\ \midrule
\multicolumn{11}{c}{\textbf{Learning-based Transpilers}}\\ \midrule
TransCoder           & 36.64\%          & 2.29\%           & 30.40\%          & 0.88\%          & 27.84\%          & 5.81\%           & 49.77\%          & 14.39\% & 36.16\% & 5.84\%         \\
TransCoder-IR        & /                & /                & /                & /               & 40.99\%          & 14.26\%          & 50.53\%          & 18.28\% & 45.76\% & 16.27\%         \\
TransCoder-ST        & 46.34\%          & 2.64\%           & 47.85\%          & 1.06\%          & 49.68\%          & 9.68\%           & 64.73\%          & 17.43\%  & 52.15\%  & 7.70\%\\ \bottomrule        
\end{tabular}
\begin{tablenotes}
\item[$\Phi$] \scriptsize For clarification, we adopt ``-'' to concatenate LLaMA and each of its corresponding parameter sizes for discrimination. For example, LLaMA with 7B parameters is dubbed LLaMA-7B.
\end{tablenotes}
\end{threeparttable}
\vspace{-1em}
\end{table}

\section{Experimental Results}
\subsection{RQ1: What is the performance of recent LLMs against state-of-the-art learning-based transpilers in code translation?}
\label{RQ1: What is the performance of recent LLMs against state-of-the-art learning-based transpilers in code translation?}
Table \ref{Empirical Results of Each Model for Code Translation} demonstrates the empirical results of each model on four translation datasets in terms of CA and EM Acc. (1) As evident from the results, irrespective of the translation dataset experimented, GPT-3.5 consistently performs the best, achieving 87.92\% in terms of CA and 18.04\% in terms of EM Acc on average among four translation datasets. Besides, LLaMA-33B also performs better than the state-of-the-art learning-based transpilers on most translation datasets in terms of CA and EM Acc. (2) In general, with the increment of parameter size, the translation performance of LLMs is improved gradually. (3) Comparing CodeGen and LLaMA-7B with similar parameter sizes but different pre-training resources and model implementations, their translation abilities are shown to be neck-to-neck in performance, where, on average, CodeGen performs better in terms of CA while LLaMA-7B obtains a higher EM Acc score. (4) It is noticeable that TransCoder-IR outperforms TransCoder-ST in terms of EM Acc on average. In contrast, TransCoder-ST obtains a higher score in terms of CA on average. A potential explanation is TransCoder-ST trained with samples filtered by unit tests, which makes it comparatively insensitive to the lexical matching but focuses more on semantic equivalency. 
\begin{boxK}
\small \faIcon{pencil-alt} \textbf{Answer to RQ1:} LLMs have achieved impressive performance in the code translation task compared with state-of-the-art transpilers, showing a promising prospect. Nonetheless, they still suffer some accuracy issues. Even the GPT-3.5 cannot carry out the perfect translation, as highlighted by the multiple failures discussed in Section \ref{Motivation}.
\end{boxK}

\begin{table}[htbp]
\vspace{-0.5em}
\setlength{\abovecaptionskip}{0.1cm}
\caption{Experimental Results of \textbf{UniTrans} with Different LLMs}
\label{Experimental Results of UniTrans with Different LLMs}
\setlength\tabcolsep{1.7pt}
\scriptsize
\begin{threeparttable}
\begin{tabular}{lcccccccccccccc}
\toprule
\multirow{2.1}{*}{\textbf{Models}}             & \multicolumn{2}{c}{\textbf{Java to Python}}                        & \multicolumn{2}{c}{\textbf{Python to Java}}                                                                       & \multicolumn{2}{c}{\textbf{C++ to Python}}                         & \multicolumn{2}{c}{\textbf{Python to C++}}                                                                         & \multicolumn{2}{c}{\textbf{Java to C++}}                           & \multicolumn{2}{c}{\textbf{C++ to Java}}                                        & \multicolumn{2}{c}{\textbf{Average}}                               \\
\cmidrule(lr){2-3} \cmidrule(lr){4-5} \cmidrule(lr){6-7} \cmidrule(lr){8-9} \cmidrule(lr){10-11}  \cmidrule(lr){12-13} \cmidrule(lr){14-15}
             & \textbf{CA}                          & \textbf{EM Acc}                      & \textbf{CA}                                                 & \textbf{EM Acc}                                              & \textbf{CA}                          & \textbf{EM Acc}                      & \textbf{CA}                                                  & \textbf{EM Acc}                                              & \textbf{CA}                          & \textbf{EM Acc}                      & \textbf{CA}                                       & \textbf{EM Acc}                      & \textbf{CA}                          & \textbf{EM Acc}                      \\ \midrule
LLaMA-7B     & 27.16\%                     & 3.35\%                      & 15.15\%                                            & 1.23\%                                              & 23.49\%                     & 3.17\%                      & 16.49\%                                             & 1.41\%                                              & 32.98\%                     & 10.04\%                     & 21.16\%                                  & 9.86\%                      & 22.74\%                     & 4.84\%                      \\
\textbf{UniTrans}     & 31.90\%                     & 3.52\%                      & 23.24\%                                            & 3.52\%                                              & 26.08\%                     & 4.23\%                      & 20.34\%                                             & 2.64\%                                              & 35.76\%                     & 13.56\%                     & 33.40\%                                  & 17.78\%                     & 28.45\%                     & 7.54\%                      \\
Improvement        & 17.45\%                     & 5.07\%                      & 53.40\%                                            & \cellcolor[HTML]{E1FFFF}\textbf{186.18\%}           & 11.03\%                     & 33.44\%                     & 23.35\%                                             & 87.23\%                                             & 8.43\%                      & 35.06\%                     & \cellcolor[HTML]{E1FFFF}\textbf{57.84\%} & 80.32\%                     & 28.58\%                     & 71.22\%                     \\ \midrule
LLaMA-13B    & 39.01\%                     & 4.58\%                      & 36.31\%                                            & 1.94\%                                              & 33.19\%                     & 4.05\%                      & 32.98\%                                             & 2.46\%                                              & 37.90\%                     & 10.39\%                     & 40.25\%                                  & 12.15\%                     & 36.61\%                     & 5.93\%                      \\
\textbf{UniTrans}     & 42.89\%                     & 4.75\%                      & 38.80\%                                            & 2.82\%                                              & 40.73\%                     & 5.28\%                      & 43.25\%                                             & 4.23\%                                              & 47.54\%                     & 12.85\%                     & 47.93\%                                  & 17.43\%                     & 43.52\%                     & 7.89\%                      \\
Improvement        & \multicolumn{1}{c}{9.95\%}  & \multicolumn{1}{c}{3.71\%}  & \multicolumn{1}{c}{6.86\%}                         & \multicolumn{1}{c}{45.36\%}                         & \multicolumn{1}{c}{22.72\%} & \multicolumn{1}{c}{30.37\%} & \multicolumn{1}{c}{\cellcolor[HTML]{E1FFFF}31.14\%} & \multicolumn{1}{c}{\cellcolor[HTML]{E1FFFF}71.95\%} & \multicolumn{1}{c}{25.44\%} & \multicolumn{1}{c}{23.68\%} & \multicolumn{1}{c}{19.08\%}              & \multicolumn{1}{c}{43.46\%} & \multicolumn{1}{c}{19.20\%} & \multicolumn{1}{c}{36.42\%} \\ \midrule
GPT-3.5       & 89.22\%                     & 10.74\%                     & 74.89\%                                            & 6.16\%                                              & 87.07\%                     & 11.44\%                     & 89.51\%                                             & 6.69\%                                              & 92.93\%                     & 27.46\%                     & 82.16\%                                  & 26.58\%                     & \multicolumn{1}{c}{85.96\%} & 14.85\%                     \\
\textbf{UniTrans}     & \multicolumn{1}{c}{91.16\%} & \multicolumn{1}{c}{11.44\%} & \multicolumn{1}{c}{81.33\%}                        & \multicolumn{1}{c}{8.10\%}                          & \multicolumn{1}{c}{88.79\%} & \multicolumn{1}{c}{12.68\%} & \multicolumn{1}{c}{94.22\%}                         & \multicolumn{1}{c}{7.57\%}                          & \multicolumn{1}{c}{94.86\%} & \multicolumn{1}{c}{29.41\%} & \multicolumn{1}{c}{85.48\%}              & \multicolumn{1}{c}{29.40\%} & \multicolumn{1}{c}{89.31\%} & 16.43\%                     \\
Improvement        & \multicolumn{1}{c}{2.17\%}  & \multicolumn{1}{c}{6.52\%}  & \multicolumn{1}{c}{\cellcolor[HTML]{E1FFFF}8.60\%} & \multicolumn{1}{c}{\cellcolor[HTML]{E1FFFF}31.49\%} & \multicolumn{1}{c}{1.98\%}  & \multicolumn{1}{c}{10.84\%} & \multicolumn{1}{c}{5.26\%}                          & \multicolumn{1}{c}{13.15\%}                         & \multicolumn{1}{c}{2.08\%}  & \multicolumn{1}{c}{7.10\%}  & \multicolumn{1}{c}{4.04\%}               & \multicolumn{1}{c}{10.57\%} & \multicolumn{1}{c}{4.02\%}  & \multicolumn{1}{c}{13.28\%} \\ \bottomrule
\end{tabular}
\begin{tablenotes}
\definecolor{light cyan}{HTML}{E1FFFF}
\fboxsep1.5pt
\item[$\Phi$] \scriptsize Table cells with \colorbox{light cyan}{light cyan} background denote the highest improvement in terms of CA/EM Acc of each LLM among various translation datasets. Digits in \textbf{bold} denote the highest improvement in terms of CA/EM Acc among all LLMs and all results.
\end{tablenotes}
\end{threeparttable}
\vspace{-2em}
\end{table}

\subsection{RQ2: How does UniTrans perform with different LLMs?}
\label{RQ2: How does UniTrans perform with different LLMs?}

Table \ref{Experimental Results of UniTrans with Different LLMs} demonstrates the experimental results of \textbf{UniTrans} embedded with various LLMs, where each row of ``Improvement'' denotes the improvement of \textbf{UniTrans} against its corresponding underlying LLM. As can be seen, \textbf{UniTrans} consistently boosts the performance of each LLM in the code translation task. Among all translation pairs, \textbf{UniTrans} brings improvements of 28.58\% in terms of CA and 71.22\% in terms of EM Acc on average for LLaMA-7B, which is the most significant among all three experimented LLMs. Particularly, the translation of Python to Java and C++ to Java respectively achieves the highest improvement in terms of EM Acc (186.18\%) and CA (57.84\%) among all results. Besides, on average, it enhances LLaMA-13B by 19.20\% and 36.42\% in terms of CA and EM Acc, where the translation of Python to C++ obtains the highest enhancement for LLaMA-13B. Moreover, it also boosts GPT-3.5 by 4.02\% and 13.28\%, on average, in terms of each evaluation metric in order, and translating from Python to Java gains the most. 

To investigate whether the improvement of \textbf{UniTrans} against each backbone is statistically significant, we perform the Wilcoxon Signed-Rank Test (WSRT) \cite{wilcoxon1992individual} with a confidence level of 95\% to make pairwise comparisons for them across six translation datasets. Subsequently, we further conduct Cliff's Delta analysis \cite{grissom2005effect} to measure the effect size in pairs. Results demonstrate that the improvements are all statistically significant and non-negligible \footnote{Detailed statistical test results are recorded in \cite{yz10191153:online}}, which further proves the superior performance of \textbf{UniTrans} in unleashing the power of LLMs in code translation. 

Although the translation between Python and Java was not extensively investigated in Section \ref{Motivation}, \textbf{UniTrans} still improves each LLM's performance on these translation datasets significantly, showing that our idea has the potential to be extended to other unseen translation pairs, and the effectiveness is also remarkable. 

\begin{boxK}
\small \faIcon{pencil-alt} \textbf{Answer to RQ2:} \textbf{UniTrans} can bring consistent and substantial improvements for different LLMs of diverse parameter sizes on various translation datasets, showing a powerful generality on various models and programming language pairs. 
\end{boxK}

\subsection{RQ3: What are the contributions of each component in UniTrans? }
\label{RQ3: What are the contributions of each component in UniTrans?}
We present Table \ref{Ablation Study of UniTrans with Different LLMs} to demonstrate the contribution of TAP and TRP of \textbf{UniTrans} on various LLMs, where 
their positive improvements in terms of CA and EM Acc are highlighted in light cyan.  
In general, adding TAP and TRP are both effective in enhancing LLMs' code translation ability, where TAP brings average improvements on CA and EM Acc by 23.77\% and 67.05\% for LLaMA-7B, by 14.67\% and 29.79\% for LLaMA-13B, and by 2.06\% and 13.02\% for GPT-3.5. Meanwhile, adding TRP enhances the metrics of CA and EM Acc on average by 3.77\% and 2.80\% for LLaMA-7B, by 4.05\% and 4.76\% for LLaMA-13B, and by 1.93\% and 0.25\% for GPT-3.5. In summary, smaller models obtain more enhancement as their room for improvement is relatively larger.

It is noticeable that GPT-3.5 suffers a little performance decline in terms of CA when using TAP for the translation from Java to Python. We manually inspect those failed cases and find most of the mistakes are tiny errors, such as mis-interchanging operands and operator misuse. We speculate that LLMs still might overlook some details even with test cases for augmentation. On the other hand, GPT-3.5 has already obtained quite good performance, leading to little room for further improvement via test cases. However, since the performance decline is very tiny (-0.48\%) and the EM Acc score here is improved (+4.93\%), we argue adding test cases as extra information is harmless for GPT-3.5 to translate code from Java to Python.  

In addition, it is also worth noting that, compared with TAP, introducing TRP brings less improvement to EM Acc, sometimes even zero. A potential explanation is that many repair processes are targeted to fix specific lines based on initial translated programs. Thus, once an initial translated program has a different implementation thought from the ground truth program, even the successfully fixed program will have difficulty matching the ground truth perfectly, causing the underestimation of the improvement. Hence, it is more important to consider the performance variation in terms of CA rather than EM Acc for TRP. Moreover, we find that TRP accounts for a larger proportion of the whole improvement in terms of CA with the increment of model size (LLaMA-7B: 3.77/23.77<LLaMA-13B:4.05/14.67<GPT-3.5:1.93/2.06). A potential explanation is larger models carry more powerful generality to program repair. Thus, even with less improvement room, they boost more than adding test cases as translation augmentation. Therefore, we stress that the TRP is also necessary for \textbf{UniTrans} and is mutually complementary to TAP.

Finally, following the procedure in RQ2, we further conduct WSRT and Cliff's Delta analysis to explore whether the improvements of TAP and TRP are significant. Results demonstrate sequentially adding TAP and TRP modules for each backbone obtains statistically significant improvements, and the effects are all non-negligible \footnote{Detailed statistical test results are recorded in \cite{yz10191153:online}. It should be noted that we did not conduct the statistical tests between TRP and TAP in terms of EM Acc because, as we analyzed in Section \ref{RQ3: What are the contributions of each component in UniTrans?}, EM Acc can hardly manifest the improvement of TRP against TAP inherently.}. 

\begin{table}[htbp]
\vspace{-0.5em}
\setlength{\abovecaptionskip}{0.1cm}
\caption{Ablation Study of \textbf{UniTrans} with Different LLMs}
\label{Ablation Study of UniTrans with Different LLMs}
\setlength\tabcolsep{2pt}
\scriptsize
\begin{threeparttable}
\begin{tabular}{lcccccccccccccc}
\toprule
\multirow{2.1}{*}{\textbf{Models}}             & \multicolumn{2}{c}{\textbf{Java to Python}}                        & \multicolumn{2}{c}{\textbf{Python to Java}}                                                                       & \multicolumn{2}{c}{\textbf{C++ to Python}}                         & \multicolumn{2}{c}{\textbf{Python to C++}}                                                                         & \multicolumn{2}{c}{\textbf{Java to C++}}                           & \multicolumn{2}{c}{\textbf{C++ to Java}}                                        & \multicolumn{2}{c}{\textbf{Average}}                               \\
\cmidrule(lr){2-3} \cmidrule(lr){4-5} \cmidrule(lr){6-7} \cmidrule(lr){8-9} \cmidrule(lr){10-11}  \cmidrule(lr){12-13} \cmidrule(lr){14-15}
             & \textbf{CA}                          & \textbf{EM Acc}                      & \textbf{CA}                                                 & \textbf{EM Acc}                                              & \textbf{CA}                          & \textbf{EM Acc}                      & \textbf{CA}                                                  & \textbf{EM Acc}                                              & \textbf{CA}                          & \textbf{EM Acc}                      & \textbf{CA}                                       & \textbf{EM Acc}                      & \textbf{CA}                          & \textbf{EM Acc}                      \\ \midrule
LLaMA-7B     & 27.16\%                     & 3.35\%                      & 15.15\%                                            & 1.23\%                                              & 23.49\%                     & 3.17\%                      & 16.49\%                                             & 1.41\%                                              & 32.98\%                     & 10.04\%                     & 21.16\%                                  & 9.86\%                      & 22.74\%                     & 4.84\%                      \\
+ TAP & 31.25\%                     & 3.35\%                      & 22.20\%                                            & 3.52\%                                              & 25.43\%                     & 4.23\%                      & 19.49\%                                             & 2.46\%                                              & 34.48\%                     & 13.38\%                     & 31.74\%                                  & 17.25\%                     & 27.43\%                     & 7.37\%                      \\
Improvement        &  \cellcolor[HTML]{E1FFFF}15.06\%                    &     0.00\%                 &  \cellcolor[HTML]{E1FFFF}46.53\%                                           & \cellcolor[HTML]{E1FFFF}186.18\%  &  \cellcolor[HTML]{E1FFFF}8.26\%                  &   \cellcolor[HTML]{E1FFFF}33.44\%                  &   \cellcolor[HTML]{E1FFFF}18.19\%                                          &                        \cellcolor[HTML]{E1FFFF}74.47\%                     &  \cellcolor[HTML]{E1FFFF}4.55\%                    &         \cellcolor[HTML]{E1FFFF}33.27\%           & \cellcolor[HTML]{E1FFFF}50.00\% &  \cellcolor[HTML]{E1FFFF}74.95\% &  \cellcolor[HTML]{E1FFFF}23.77\%                   & \cellcolor[HTML]{E1FFFF}67.05\%                     \\
+ TRP     & 31.90\%                     & 3.52\%                      & 23.24\%                                            & 3.52\%                                              & 26.08\%                     & 4.23\%                      & 20.34\%                                             & 2.64\%                                              & 35.76\%                     & 13.56\%                     & 33.40\%                                  & 17.78\%                     & 28.45\%                     & 7.54\%                      \\
Improvement        &  \cellcolor[HTML]{E1FFFF}2.08\%                    &     \cellcolor[HTML]{E1FFFF}5.07\%                 &   \cellcolor[HTML]{E1FFFF}4.68\%                                        &   0.00\%  &  \cellcolor[HTML]{E1FFFF}2.56\%                  &   0.00\%                  &      \cellcolor[HTML]{E1FFFF}4.36\%                                       &    \cellcolor[HTML]{E1FFFF}7.32\%                                         &   \cellcolor[HTML]{E1FFFF}3.71\%                   &   \cellcolor[HTML]{E1FFFF}1.35\%                 & \cellcolor[HTML]{E1FFFF}5.23\% & \cellcolor[HTML]{E1FFFF}3.07\%  &    \cellcolor[HTML]{E1FFFF}3.77\%                 &   \cellcolor[HTML]{E1FFFF}2.80\%                   \\ \midrule
LLaMA-13B    & 39.01\%                     & 4.58\%                      & 36.31\%                                            & 1.94\%                                              & 33.19\%                     & 4.05\%                      & 32.98\%                                             & 2.46\%                                              & 37.90\%                     & 10.39\%                     & 40.25\%                                  & 12.15\%                     & 36.61\%                     & 5.93\%                      \\
+ TAP & 42.03\%                     & 4.75\%                      & 36.72\%                                            & 2.64\%                                              & 39.44\%                     & 4.93\%                      & 42.18\%                                             & 3.87\%                                              & 46.47\%                     & 12.85\%                     & 44.19\%                                  & 16.55\%                     & 41.84\%                     & 7.60\%                      \\
Improvement        &  \cellcolor[HTML]{E1FFFF}7.74\%                    &            \cellcolor[HTML]{E1FFFF}3.71\%          &    \cellcolor[HTML]{E1FFFF}1.13\%                                         & \cellcolor[HTML]{E1FFFF}36.08\%  &    \cellcolor[HTML]{E1FFFF}18.83\%                &    \cellcolor[HTML]{E1FFFF}21.73\%                 &         \cellcolor[HTML]{E1FFFF}27.90\%                                    &    \cellcolor[HTML]{E1FFFF}57.32\%                                         &    \cellcolor[HTML]{E1FFFF}22.61\%                  &  \cellcolor[HTML]{E1FFFF}23.68\%                  & \cellcolor[HTML]{E1FFFF}9.79\% &  \cellcolor[HTML]{E1FFFF}36.21\% &   \cellcolor[HTML]{E1FFFF}14.67\%                  &        \cellcolor[HTML]{E1FFFF}29.79\%              \\
+ TRP     & 42.89\%                     & 4.75\%                      & 38.80\%                                            & 2.82\%                                              & 40.73\%                     & 5.28\%                      & 43.25\%                                             & 4.23\%                                              & 47.54\%                     & 12.85\%                     & 47.93\%                                  & 17.43\%                     & 43.52\%                     & 7.89\%                      \\
Improvement        & \cellcolor[HTML]{E1FFFF}2.05\%                     &               0.00\%       &     \cellcolor[HTML]{E1FFFF}5.66\%                                        &  \cellcolor[HTML]{E1FFFF}6.82\% &  \cellcolor[HTML]{E1FFFF}3.27\%                  &     \cellcolor[HTML]{E1FFFF}7.10\%                &   \cellcolor[HTML]{E1FFFF}2.54\%                                          & \cellcolor[HTML]{E1FFFF}9.30\%                                            &   \cellcolor[HTML]{E1FFFF}2.30\%                   &   0.00\%                 & \cellcolor[HTML]{E1FFFF}8.46\% & \cellcolor[HTML]{E1FFFF}5.32\%  &  \cellcolor[HTML]{E1FFFF}4.05\%                   &    \cellcolor[HTML]{E1FFFF}4.76\%                  \\ \midrule
GPT-3.5       & 89.22\%                     & 10.74\%                     & 74.89\%                                            & 6.16\%                                              & 87.07\%                     & 11.44\%                     & 89.51\%                                             & 6.69\%                                              & 92.93\%                     & 27.46\%                     & 82.16\%                                  & 26.58\%                     & \multicolumn{1}{c}{85.96\%} & 14.85\%                     \\
+ TAP & 88.79\%                     & 11.27\%                     & 79.67\%                                            & 8.10\%                                              & 87.28\%                     & 12.68\%                     & 92.72\%                                             & 7.57\%                                              & 94.43\%                     & 29.41\%                     & 82.99\%                                  & 29.39\%                     & 87.65\%                     & 16.40\%                     \\
Improvement        & -0.48\%                     &            \cellcolor[HTML]{E1FFFF}4.93\%          &                                    \cellcolor[HTML]{E1FFFF}6.38\%         & \cellcolor[HTML]{E1FFFF}31.49\%  &   \cellcolor[HTML]{E1FFFF}0.24\%                &   \cellcolor[HTML]{E1FFFF}10.84\%                  &                                      \cellcolor[HTML]{E1FFFF}3.59\%       &        \cellcolor[HTML]{E1FFFF}13.15\%                                     &          \cellcolor[HTML]{E1FFFF}1.61\%             &    \cellcolor[HTML]{E1FFFF}7.10\%                & \cellcolor[HTML]{E1FFFF}1.01\% & \cellcolor[HTML]{E1FFFF}10.57\%  &                 \cellcolor[HTML]{E1FFFF}2.06\%    &    \cellcolor[HTML]{E1FFFF}13.02\%                  \\
+ TRP     & \multicolumn{1}{c}{91.16\%} & \multicolumn{1}{c}{11.44\%} & \multicolumn{1}{c}{81.33\%}                        & \multicolumn{1}{c}{8.10\%}                          & \multicolumn{1}{c}{88.79\%} & \multicolumn{1}{c}{12.68\%} & \multicolumn{1}{c}{94.22\%}                         & \multicolumn{1}{c}{7.57\%}                          & \multicolumn{1}{c}{94.86\%} & \multicolumn{1}{c}{29.41\%} & \multicolumn{1}{c}{85.48\%}              & \multicolumn{1}{c}{29.40\%} & \multicolumn{1}{c}{89.31\%} & 16.43\%                     \\
Improvement        &  \cellcolor[HTML]{E1FFFF}2.67\%                    &      \cellcolor[HTML]{E1FFFF}1.51\%                &   \cellcolor[HTML]{E1FFFF}2.08\%                                          & 0.00\%  &  \cellcolor[HTML]{E1FFFF}1.73\%                  & 0.00\%                    &  \cellcolor[HTML]{E1FFFF}1.62\%                                           &   0.00\%                                          &  \cellcolor[HTML]{E1FFFF}0.46\%                    &   0.00\%                 & \cellcolor[HTML]{E1FFFF}3.00\% &  0.00\% &  \cellcolor[HTML]{E1FFFF}1.93\%                   &  \cellcolor[HTML]{E1FFFF}0.25\%                    \\ \bottomrule
\end{tabular}
\begin{tablenotes}
\item[$\Phi$] \scriptsize TAP and TRP are the abbreviations of the Translation Augmentation Phase and Translation Repair Phase, which is the same as Table \ref{Test Case Number Analysis for UniTrans with LLaMA-7B/GPT-3.5}, \ref{Iterative Repair Analysis for UniTrans with LLaMA-7B/GPT-3.5} and \ref{The Efficacy of UniTrans on Various Categories of Mistakes, Taking GPT-3.5 as An Example}.
\end{tablenotes}
\end{threeparttable}
\vspace{-1em}
\end{table}

\begin{boxK}
\small \faIcon{pencil-alt} \textbf{Answer to RQ3:} Each component, i.e., (1) the Translation Augmentation Phase and (2) the Translation Repair Phase, is essential for \textbf{UniTrans}. Three LLMs have been progressively improved with the addition of each component gradually.
\end{boxK}

\begin{table}[htbp]
\setlength{\abovecaptionskip}{0.1cm}
\caption{Test Case Number Analysis for \textbf{UniTrans} with LLaMA-7B/GPT-3.5}
\label{Test Case Number Analysis for UniTrans with LLaMA-7B/GPT-3.5}
\scriptsize
\begin{threeparttable}
\begin{tabular}{cccccccccc}
\toprule
\multirow{2.1}{*}{\textbf{\# Test Cases}} & \multirow{2.1}{*}{\textbf{Line Coverage}}               & \multicolumn{2}{c}{\textbf{Python to Java (TAP)}}                          & \multicolumn{2}{c}{\textbf{Python to Java (TRP)}}                          & \multicolumn{2}{c}{\textbf{C++ to Java (TAP)}}                              & \multicolumn{2}{c}{\textbf{C++ to Java (TRP)}}                              \\ \cmidrule(lr){3-4} \cmidrule(lr){5-6} \cmidrule(lr){7-8} \cmidrule(lr){9-10}
                       &                                      & \textbf{CA}                          & \textbf{EM Acc}                     & \textbf{CA}                          & \textbf{EM Acc}                     & \textbf{CA}                          & \textbf{EM Acc}                      & \textbf{CA}                          & \textbf{EM Acc}                      \\ \midrule
\multicolumn{10}{c}{\textbf{UniTrans with LLaMA-7B}}                                \\ \midrule
0                      & 0                                    & 15.15\%                              & 1.23\%                              & /                                    & /                                   & 21.16\%                              & 9.86\%                               & /                                  & /                                    \\
1                      & 78.84\%                              & 21.58\%                              & \textbf{3.52\%}                     & 23.03\%                              & \textbf{3.70\%}                     & 29.88\%                              & \textbf{18.13\%}                     & 30.91\%                              & \textbf{18.49\%}                     \\
3                      & 85.20\%                              & \textbf{22.20\%}                     & \textbf{3.52\%}                     & \textbf{23.24\%}                     & 3.52\%                              & \textbf{31.74\%}                     & 17.25\%                              & \textbf{33.40\%}                     & 17.78\%                              \\
5                      & \textbf{87.20\%}                     & 20.33\%                              & 2.99\%                              & 21.58\%                              & 2.99\%                              & 27.59\%                              & 16.20\%                              & 29.67\%                              & 17.43\%                              \\ \midrule
\multicolumn{10}{c}{\textbf{UniTrans with GPT-3.5}}                                   \\ \midrule
0                      & 0                      & 74.89\%                  & 6.16\%                 & /                        & /                      & 82.16\%                & 26.58\%               & /                      & /                     \\
1                      & 82.57\%                & 79.25\%                  & 7.92\%                 & 80.91\%                  & 7.80\%                 & 82.34\%                & 25.53\%               & 84.54\%                & 25.63\%               \\
3                      & 92.14\%                & \textbf{79.67\%}         & 8.10\%                 & \textbf{81.33\%}         & 8.10\%                 & \textbf{82.99\%}       & \textbf{29.39\%}      & \textbf{85.48\%}       & \textbf{29.40\%}      \\
5                      & \textbf{94.50\%}       & 76.35\%                  & \textbf{8.63\%}        & 79.38\%                  & \textbf{8.45\%}        & 79.88\%                & 28.35\%               & 82.76\%                & 28.87\% \\ \bottomrule                  
\end{tabular}
\begin{tablenotes}
\item[$\Phi$] \scriptsize We label the translation pair with ``(TAP)'' or ``(TRP)'' to indicate their performance after each phase. Since it is impossible to identify error programs without test cases, we do not apply TRP for trials with zero test cases.
\end{tablenotes}
\end{threeparttable}
\vspace{-1em}
\end{table}

\subsection{RQ4: How does the number of test cases influence the performance of UniTrans?}

Table \ref{Test Case Number Analysis for UniTrans with LLaMA-7B/GPT-3.5} demonstrates the experimental results of test case number analysis on the TAP and TRP of \textbf{UniTrans}, respectively. Initially, with the increment of the test case sampling budget, line coverage rates gradually increase, and the performance of LLMs is also progressively enhanced, which means more test cases can provide a broader vision for LLMs to comprehend the code. However, the improvements in both models do not last long. LLaMA-7B achieves its best when given test cases no more than 3, while GPT-3.5 can further improve its performance in some situations with 5 test cases.
A potential explanation for this variation is even though more test cases offer higher code coverage rates and more chances to learn the code, the marginal returns are gradually decreasing. In this case, apart from the benefits mentioned above, excessive test cases likewise introduce more redundancies, leading to both backbone models being distracted to different extents in the Translation Augmentation Phase, which further hinders the improvement of the Translation Repair Phase.

\begin{boxK}
\small \faIcon{pencil-alt} \textbf{Answer to RQ4:} Incorporating more test cases indeed offers a more extensive vision to LLMs to translate code, but it also brings redundancies and distracts LLMs. Based on the experimental results above, we recommend larger LLMs can be fed with relatively more test cases, while smaller LLMs are suggested to offer fewer test cases.
\end{boxK}

\begin{table}[htbp]
\vspace{-1em}
\setlength{\abovecaptionskip}{0.1cm}
\caption{Valid Input Generation Analysis for \textbf{UniTrans} with LLaMA-7B/GPT-3.5}
\label{Valid Input Generation Analysis for UniTrans with LLaMA-7B/GPT-3.5}
\setlength\tabcolsep{3pt}
\scriptsize
\begin{threeparttable}
\begin{tabular}{ccccccccccccc}
\toprule
\multirow{2.2}{*}{\textbf{\# Valid Inputs}} & \multicolumn{3}{c}{\textbf{Python}}                    & \multicolumn{3}{c}{\textbf{Java}}                      & \multicolumn{3}{c}{\textbf{C++}}                       & \multicolumn{3}{c}{\textbf{Average}}                   \\ \cmidrule(lr){2-4} \cmidrule(lr){5-7} 
\cmidrule(lr){8-10}  \cmidrule(lr){11-13}
\textbf{}      & \textbf{1}       & \textbf{2}       & \textbf{3}       & \textbf{1}       & \textbf{2}       & \textbf{3}       & \textbf{1}       & \textbf{2}       & \textbf{3}       & \textbf{1}       & \textbf{2}       & \textbf{3}       \\ \midrule
\multicolumn{13}{c}{\textbf{UniTrans with LLaMA-7B}}      \\ \midrule
1              & \textit{70.42\%}          & 84.51\%          & 95.07\%          & 78.52\%          & 89.79\%          & 97.01\%          & 69.37\%          & \textbf{84.68\%} & 95.77\%          & 72.77\%          & 86.33\%          & 93.84\%          \\
3              & 66.55\%          & 82.39\%          & 92.78\%          & 73.24\%          & 87.15\%          & 95.95\%          & 63.73\%          & 79.75\%          & 92.08\%          & 67.84\%          & 83.10\%          & 92.31\%          \\
5              & 63.38\%          & 79.40\%          & 91.55\%          & 68.13\%          & 85.04\%          & 94.19\%          & 59.15\%          & 74.82\%          & 88.38\%          & 63.55\%          & 79.75\%          & 91.26\%          \\ \midrule
\multicolumn{13}{c}{\textbf{UniTrans with GPT-3.5}}        \\ \midrule
1              & \textbf{96.48\%} & \textbf{97.01\%} & \textbf{97.36\%} & \textbf{96.83\%} & \textbf{97.18\%} & \textbf{97.18\%} & \textbf{78.52\%} & 83.80\%          & \textbf{89.44\%} & \textbf{90.61\%} & \textbf{92.66\%} & \textbf{96.77\%} \\
3              & \textbf{94.89\%} & \textbf{97.01\%} & \textbf{97.18\%} & \textbf{96.30\%} & \textbf{97.01\%} & \textbf{97.18\%} & \textbf{78.52\%} & \textbf{83.80\%} & \textbf{88.20\%} & \textbf{89.90\%} & \textbf{92.61\%} & \textbf{95.48\%} \\
5              & \textbf{94.36\%} & \textbf{96.48\%} & \textbf{97.01\%} & \textbf{95.95\%} & \textbf{97.01\%} & \textbf{97.01\%} & \textbf{78.52\%} & \textbf{83.80\%} & \textbf{88.03\%} & \textbf{89.61\%} & \textbf{92.43\%} & \textbf{94.13\%} \\ \bottomrule
\end{tabular}
\begin{tablenotes}
\item[$\Phi$] \scriptsize \# Valid Inputs denotes the number of valid inputs measured per experiment, where each experiment carries out $\{1,2,3\}$ attempts of inference.
\end{tablenotes}
\end{threeparttable}
\vspace{-2em}
\end{table}

\subsection{RQ5: What is the performance of LLMs on the valid input generation task?}

Table \ref{Valid Input Generation Analysis for UniTrans with LLaMA-7B/GPT-3.5} records the rate of generating at least $\{1, 3, 5\}$ valid inputs given $\{1, 2, 3\}$ attempts of inference. For example, \textit{70.42\%} (highlighted in italics in Table \ref{Valid Input Generation Analysis for UniTrans with LLaMA-7B/GPT-3.5}) denotes LLaMA-7B can generate at least one valid input for 70.42\% of programs given one attempt of inference. Apparently, GPT-3.5 outperforms LLaMA-7B in most of the settings, as it carries more parameters and a more powerful generality. In the meantime, when provided with only three attempts of inference for each of the two models, GPT-3.5 is capable of generating at least 1/3/5 valid inputs for an average of 96.77\%/95.48\%/94.13\% of programs, while LLaMA-7B achieves this for 93.84\%/92.31\%/91.26\%\ of programs on average, showing that the LLMs' capabilities of valid input generation for programs are highly efficient. As obtaining valid inputs is equivalent to obtaining test cases, the above results further demonstrate the high efficiency of our proposed Test Case Generation Phase, proving the practicability of \textbf{UniTrans} for future deployment.

\begin{boxK}
\small \faIcon{pencil-alt} \textbf{Answer to RQ5:} LLMs with various parameter sizes are capable of generating valid inputs for over 90\% programs with few attempts, showing that \textbf{UniTrans}, a code translation framework that is highly dependent on test cases, is feasible and efficient in practice.  
\end{boxK}

\begin{table}[htbp]
\vspace{-1em}
\setlength{\abovecaptionskip}{0.1cm}
\caption{Iterative Repair Analysis for \textbf{UniTrans} with LLaMA-7B/GPT-3.5}
\label{Iterative Repair Analysis for UniTrans with LLaMA-7B/GPT-3.5}
\setlength\tabcolsep{2.5pt}
\scriptsize
\begin{threeparttable}
\begin{tabular}{ccccccccccccccc}
\toprule
\multirow{2.2}{*}{\textbf{\# Max Iter.}}           & \multicolumn{2}{c}{\textbf{Java to Python}} & \multicolumn{2}{c}{\textbf{Python to Java}}  & \multicolumn{2}{c}{\textbf{C++ to Python}}   & \multicolumn{2}{c}{\textbf{Python to C++}} & \multicolumn{2}{c}{\textbf{Java to C++}} & \multicolumn{2}{c}{\textbf{C++ to Java}}     & \multicolumn{2}{c}{\textbf{Average}} \\ \cmidrule(lr){2-3} \cmidrule(lr){4-5} \cmidrule(lr){6-7} \cmidrule(lr){8-9}
\cmidrule(lr){10-11} \cmidrule(lr){12-13} \cmidrule(lr){14-15}           & \textbf{CA}                   & \textbf{PR}          & \textbf{CA}               & \textbf{PR}               & \textbf{CA}               & \textbf{PR}               & \textbf{CA}                   & \textbf{PR}         & \textbf{CA}                  & \textbf{PR}        & \textbf{CA}               & \textbf{PR}               & \textbf{CA}                & \textbf{PR}      \\ \midrule
\multicolumn{15}{c}{\textbf{UniTrans with LLaMA-7B}}  \\ \midrule
+ Repair-0 & 31.25\%              & 38.66\%     & 22.20\%          & 25.81\%          & 25.43\%          & 33.23\%          & 19.49\%              & 22.31\%    & 34.48\%             & 36.02\%   & 31.74\%          & 32.12\%          & 27.43\%           & 31.36\% \\
+ Repair-1 & 31.90\%              & 39.63\%     & 23.24\%          & 28.15\%          & 26.08\%          & 34.07\%          & 20.34\%              & 23.47\%    & 35.76\%             & 37.32\%   & 33.40\%          & 33.92\%          & 28.45\%           & 32.76\% \\
+ Repair-2 & \cellcolor[HTML]{E1FFFF}\textbf{32.11\%} & \cellcolor[HTML]{E1FFFF}39.94\%     & 23.65\%          & 28.55\%          & 26.29\%          & 34.14\%          & \cellcolor[HTML]{E1FFFF}\textbf{20.56\%} & \cellcolor[HTML]{E1FFFF}23.92\%    & 35.97\%             & 37.41\%   & 33.82\%          & 34.69\%          & 28.73\%  & 33.11\% \\
+ Repair-3 & 31.82\%              & 39.39\%     & \cellcolor[HTML]{E1FFFF}\textbf{24.38\%} & \cellcolor[HTML]{E1FFFF}29.14\%          & \cellcolor[HTML]{E1FFFF}\textbf{26.94\%} & \cellcolor[HTML]{E1FFFF}34.72\%          & 20.12\%              & 23.79\%    & \cellcolor[HTML]{E1FFFF}\textbf{36.19\%}    & \cellcolor[HTML]{E1FFFF}37.79\%   & \cellcolor[HTML]{E1FFFF}\textbf{34.02\%} & \cellcolor[HTML]{E1FFFF}34.90\%          & \cellcolor[HTML]{E1FFFF}\textbf{28.91\%}           & \cellcolor[HTML]{E1FFFF}33.29\% \\ \midrule
\multicolumn{15}{c}{\textbf{UniTrans with GPT-3.5}}   \\ \midrule
+ Repair-0 & 88.79\%              & 91.87\%     & 79.67\%          & 81.84\%          & 87.28\%          & 91.12\%          & 92.72\%              & 95.76\%    & 94.43\%             & 95.89\%   & 82.99\%          & 83.88\%          & 87.65\%           & 90.06\% \\
+ Repair-1 & 91.16\%              & 93.73\%     & 81.33\%          & 82.93\%          & 88.79\%          & 92.41\%          & 94.22\%              & 96.81\%    & 94.86\%             & 95.87\%   & 85.48\%          & 86.51\%          & 89.31\%           & 91.38\% \\
+ Repair-2 & \cellcolor[HTML]{E1FFFF}91.16\%     & \cellcolor[HTML]{E1FFFF}\textbf{93.77\%}     & \cellcolor[HTML]{E1FFFF}81.74\%          & \cellcolor[HTML]{E1FFFF}\textbf{83.88\%} & 89.66\%          & 93.00\%          & \cellcolor[HTML]{E1FFFF}\textbf{94.65\%}     & \cellcolor[HTML]{E1FFFF}96.92\%    & 95.72\%             & 96.83\%   & \cellcolor[HTML]{E1FFFF}85.89\%          & \cellcolor[HTML]{E1FFFF}\textbf{86.72\%} & \cellcolor[HTML]{E1FFFF}\textbf{89.80\%}  & \cellcolor[HTML]{E1FFFF}91.85\% \\
+ Repair-3 & 90.95\%              & 93.51\%     & 81.74\%          & 83.80\%          & \cellcolor[HTML]{E1FFFF}89.66\%          & \cellcolor[HTML]{E1FFFF}\textbf{93.02\%} & 94.43\%              & 96.92\%    & \cellcolor[HTML]{E1FFFF}\textbf{95.93\%}    & \cellcolor[HTML]{E1FFFF}96.40\%   & 85.89\%          & 86.54\%          & 89.77\%           & 91.70\% \\ \bottomrule
\end{tabular}
\begin{tablenotes}
\item[$\Phi$] \scriptsize + Repair-$\{1,2,3\}$ denotes iteratively evaluating the repaired programs and repeating the TRP for $\{1,2,3\}$ times in \textbf{UniTrans}. In particular, + Repair-0 denotes no TRP when constructing \textbf{UniTrans}. 
\end{tablenotes}
\end{threeparttable}
\vspace{-1.0em}
\end{table}

\subsection{RQ6: How do the rounds of iterative repair affect the performance of UniTrans?}

Table \ref{Iterative Repair Analysis for UniTrans with LLaMA-7B/GPT-3.5} presents the experimental results of \textbf{UniTrans} with LLaMA-7B/GPT-3.5 under various maximum repair iterations in terms of CA and PR. For both LLaMA-7B and GPT-3.5, we highlight their best-performing repair iteration for each translation pair in light cyan, while the decisive metrics are highlighted in bold, following the evaluation criterion presented in Section \ref{Research Questions and Evaluation Methodology}. As can be seen, with more repair attempts, regardless of LLaMA-7B or GPT-3.5, the marginal returns of the performance decline gradually. In most situations, GPT-3.5 achieves its best performance when repeating the repair 2 times, while LLaMA-7B can further improve its performance even if the repair has been repeated 3 times, which is also consistent with their average performance. The explanation is intuitive, as GPT-3.5, with a way larger parameter size, carries a much more powerful generality to program repair than LLaMA-7B, leading it to fix the bugs with fewer steps. On the other hand, The remaining buggy code for GPT-3.5 is less than LLaMA-7B. Thus, there is also less room for GPT-3.5 to further improve. In addition, \textbf{UniTrans} only repairs those true error programs as they are filtered out by auto-generated test cases. Thus, their performance should be no change at worst in terms of CA in theory, whereas some of their performance, in fact, declines a little bit with more rounds of repair, such as repairing in the 3rd round with LLaMA-7B/GPT-3.5 for the translation dataset of Python to C++. We manually examine those affected samples and come to our conclusion: some error programs cannot be revealed by the evaluation-purpose test suite but are identified by our auto-generated test cases. However, \textbf{UniTrans} cannot correctly fix them. In contrast, it makes them further away from the ground truths, leading to performance declines based on this set of evaluation metrics.   
\begin{boxK}
\small \faIcon{pencil-alt} \textbf{Answer to RQ6:} Iteratively repairing error programs can fix more bugs but also bring the risk of degeneration. We suggest practitioners using \textbf{UniTrans} with larger LLMs to perform fewer rounds of repair, while for smaller LLMs, we recommend performing repair for more rounds.  
\end{boxK}

\begin{table}[htbp]
\vspace{-0.5em}
\setlength{\abovecaptionskip}{0.1cm}
\caption{The Efficacy of \textbf{UniTrans} on Various Categories of Failures, Taking GPT-3.5 as An Example}
\label{The Efficacy of UniTrans on Various Categories of Mistakes, Taking GPT-3.5 as An Example}
\definecolor{light cyan}{HTML}{E1FFFF}
\fboxsep1.5pt
\setlength\tabcolsep{2.5pt}
\scriptsize
\begin{threeparttable}
\begin{tabular}{lcccccc}
\toprule
\textbf{Statistics}                         & \textbf{Logic}                       & \textbf{Syntax}                               & \textbf{I/O}          & \textbf{API} & \textbf{Precision} & \textbf{Others} \\ \midrule
Total Failures           & 67                                   & 38                                            & 26                    & 14           & 20                 & 9               \\
Improvement by TAP       & \colorbox{light cyan}{23} (34.32\%) & 17 (44.74\%)                                  & 12 \textbf{(46.15\%)} & 5 (35.71\%)  & 3 (15\%)           & 4 (44.44\%)     \\
Total Failures after TAP & 44                                   & 21                                            & 14                    & 9            & 17                 & 5               \\
Improvement by TRP       & 8 (18.18\%)                          & \colorbox{light cyan}{10} \textbf{(47.62\%)} & 4 (28.57\%)           & 1 (11.11\%)  & 2 (11.76\%)        & 2 (40\%)        \\
Total Improvement        & \colorbox{light cyan}{31} (46.27\%) & 27 \textbf{(71.05\%)}                         & 16 (61.54\%)          & 6 (42.86\%)  & 5 (25\%)           & 6 (66.67\%)   \\ \bottomrule
\end{tabular}
\begin{tablenotes}
\item[$\Phi$] \scriptsize Digits in \colorbox{light cyan}{light cyan} and \textbf{bold} denote the most failure types that TAP/TRP addresses in quantity and proportion, respectively.  
\end{tablenotes}
\end{threeparttable}
\vspace{-1em}
\end{table}

\section{Discussion}
\subsection{Case Study}
In this section, we list two examples to qualitatively present the advantage of adding test cases for augmenting code translation and repairing error programs. (1) Figure \ref{Basic Prompt} demonstrates a failed case of using \textit{Basic Prompt}, where GPT-3.5 cannot fully capture the logic of the source program and make a Logic failure. In contrast, when we add several test cases for augmentation, as shown in Figure \ref{An Example of Translation Augmentation Prompt}, GPT-3.5 rectifies its previous mistake and completes the translation correctly. (2) Another example can be found in Figure \ref{An Example of Repair Prompt}, a code translation sample has been shipped to the Translation Repair Phase due to its failure in the Translation Augmentation Phase. As can be seen, although given several test cases for translation augmentation, GPT-3.5 still cannot correctly translate the source program to the target program due to the Java API misuse, i.e., Java uses ``length()'' to get lengths of strings, rather than ``size()''. By identifying the buggy lines and pointing out the error message, GPT-3.5 successfully repairs the error program.

\subsection{Failure Taxonomy Look Back}
As mentioned in Section \ref{Motivation}, we partition LLMs' failures into six categories, taking GPT-3.5 as an example. This section looks back at the failure taxonomy and discusses the enhancement of \textbf{UniTrans} against GPT-3.5 on each failure type so as to summarize the future direction. As seen from the results in Table \ref{The Efficacy of UniTrans on Various Categories of Mistakes, Taking GPT-3.5 as An Example}, TAP solves 46.15\% of I/O failures and 23 Logic failures, which are the highest in terms of proportion and quantity, respectively, among various failures. It demonstrates adding test cases with I/O types for translation augmentation is effective for LLMs' to comprehend program logic and the requirement on I/O. Moreover, TAP also has a commendable ability to resolve other mistakes, such as Syntax (44.74\%) and API (35.71\%).
For the remaining mistakes, TRP further repairs them based on the translated programs in TAP. Apparently, it exhibits the most superior ability in handling Syntax failures, where 10 such failures are addressed, accounting for 47.62\% of the various types of mistakes. In general, both TAP and TRP are capable of resolving those mistakes as we expected, but they still fail in certain cases, especially on Precision failures (only 25\% are solved in total). We speculate that these mistakes are so tiny that they result in only minor discrepancies with the ground truth programs, which would be extremely imperceptible to an LLM. As such, we emphasize that one of the future directions for code translation with LLMs is how to make LLMs perceive those minor discrepancies, thereby further improving the reliability of automated code translation.

\subsection{Threats to Validity}
In this section, we categorize threats to the validity of this work from three aspects.

\textbf{Internal Validity:} One threat to the internal validity comes from the potential data leakage of LLMs, which means there may be some overlaps between the training set of LLMs and the testing set used in this work. However, according to our compared learning-based transpilers \cite{roziere2020unsupervised,roziere2021leveraging,szafraniec2022code} and open-source LLMs \cite{touvron2023llama,nijkamp2022codegen}, although those LLMs were also pre-trained on natural languages, as for training resources on code, all the above models were pre-trained based on GitHub public dataset available on Google BigQuery\footnote{https://console.cloud.google.com/marketplace/details/github/github-repos}. However, the evaluation dataset crafted by \cite{roziere2020unsupervised,roziere2021leveraging,szafraniec2022code} was extracted from an independent online platform, namely GeeksforGeeks \cite{Geeksfor48:online}, and \cite{roziere2020unsupervised,roziere2021leveraging,szafraniec2022code} claimed that there is no data overlap between the pre-training and evaluation datasets.
As for GPT-3.5, a closed-source LLM, it is hard to know their pre-training resources. An alternative way to inspect the overlap between the training and testing set is to evaluate GPT-3.5 on a new testing set with minimal leakage possibility and check the performance variation between the original and new testing sets. To do this, we curated 82 C++ codes manually written by some undergraduate students. These codes were submitted via a university's private Online Judge (OJ) platform from Nov. 2023 to Jan. 2024. Thus, the OJ platform is invisible from outside the university. Besides, GPT-3.5-turbo-0613 (we used version) is a snapshot before June 13th, 2023. Therefore, it cannot exploit codes written after its release date for training. In summary, this evaluation dataset's possibility of data leakage is minimal. Afterward, we leverage GPT-3.5 to translate our collected 82 C++ code to Java and Python with a one-shot setting (same as the default setting in our manuscript), respectively. As the OJ platform already has the Java and Python test suites for evaluation but without the ground truth for comparison, we only assess GPT-3.5 in terms of CA, which is 81.71\% on C++ to Python translation and 80.49\% on C++ to Java translation. Compared to the experimental results in Section \ref{RQ1: What is the performance of recent LLMs against state-of-the-art learning-based transpilers in code translation?}, the performance drops 6.16\% on C++ to Python while drops 2.03\% on C++ to Java. Considering the code snippets of the newly crafted dataset are longer than the C++ dataset used in our manuscript (17.23 v.s. 12.94), the performance's slight degeneration is acceptable. Hence, we conclude that the threat to the training/testing overlap is very limited. All evaluation results and datasets used in this section can be found in \cite{yz10191153:online}.  


Besides, the replication of the baselines and evaluation metrics are also two threats to the internal validity. To minimize these threats, we strictly followed their replication documents and directly utilized their source code for model construction. All baselines are consistently evaluated on our cleaned dataset. As for the replication of evaluation metrics, e.g., CA, EM Acc, and PR, we reused the published code from \cite{roziere2020unsupervised,10.1109/ICSE48619.2023.00179} to implement them. 

\textbf{External Validity:} The first threat to the external validity lies in the quality of the evaluation dataset, which was released by \cite{roziere2020unsupervised}, and has been widely used in the code translation field owing to its multi-lingual samples and self-contained test suites. However, the dataset is flawed, as we mentioned in Section \ref{Data Cleaning}, and manually cleaning the dataset may not eliminate all noises or errors. To address this, we required the four authors to cross-check their cleaned results in pairs to reach a consensus for each group of parallel functions. Furthermore, we release the cleaned dataset for public evaluation. 

The limited choice of experimental models is another threat. In this work, we experimented with the most widely examined LLMs in different families and sizes \cite{yang2024improving,chen2022codet,li2023towards,xia2023automated}. For example, we select open-source LLaMA-7B/13B/33B to make a comparison among LLMs of the same family but different sizes, and we also include CodeGen-6B to compare with LLaMA-7B to explore the influence between different LLM families. Furthermore, we introduce GPT-3.5 for experiments because it is a typical closed-source LLM. More importantly, all the above selected LLMs can generate code given specific requirements, ensuring the accomplishment of the experiment. Besides, we select TransCoder, TransCoder-IR, and TransCoder-ST for comparison, as they are the state-of-the-art transpilers to date, ensuring an in-depth investigation and analysis. Therefore, we believe this threat is minimal and will include more relevant models for experiments in the future.

\textbf{Construct Validity:} The property of evaluation metrics is a primary threat \cite{yang2021multi,yang2023significance,zhang2022improving}. In this work, we adopt CA and EM Acc to evaluate the code translation performance of different approaches from both lexical and semantic correctness as most of the relevant papers \cite{roziere2020unsupervised,roziere2021leveraging,szafraniec2022code,10.1109/ICSE48619.2023.00072}. Besides, we also include PR to conduct the measurement in a more fine-grained manner \cite{10.1109/ICSE48619.2023.00179}. Therefore, we believe the evaluation is comprehensive.

\section{Conclusion}
This work aims to enhance the efficiency and reliability of codebase migration. To achieve this, we propose to leverage LLMs to substitute learning-based transpilers. Hence, we first conduct an extensive empirical study and an in-depth analysis to investigate the strengths and weaknesses of recent LLMs compared with current transpilers. Enlightened by our findings in the empirical study, we design a \textbf{Uni}fied code \textbf{Trans}lation framework for diverse LLMs, namely \textbf{UniTrans}, which incorporates test cases to augment code translation and repair bugs for incorrectly translated programs. Comprehensive experiments are carried out and demonstrate \textbf{Unitrans} improves various LLMs' capabilities in code translation by a significant margin. Below, we summarize the implications of this work for practitioners and researchers, respectively.

\textbf{Implications for practitioners:} \textbf{UniTrans} is an automated code translation framework that can effectively boost the code translation performance of various LLMs, as substantiated in Section \ref{RQ2: How does UniTrans perform with different LLMs?}. It exempts large-scale fine-tuning from LLMs and only needs to use a series of well-designed prompts and the execution of auto-generated test cases to effectively improve the performance of LLMs. Notably, Practitioners are only required to designate a handful of hyperparameters, such as the number of test cases in TAP and the round of repair in TRP, then \textbf{UniTrans} can be encapsulated into a complete system for codebase migration, showing its high practical utility and aligning with contemporary demands for efficiency and resource optimization in code translation processes. 

\textbf{Implications for researchers:} This work empirically explored the prospects and limitations of recent LLMs in automated code translation. Subsequently, we conducted the inaugural failure taxonomy grounded in the translation results of the best-performing LLM and summarized a series of improvement directions. Researchers can further propose more effective methods to unleash the power of LLMs or even re-construct them based on our insightful findings. While the prompts meticulously devised in each phase of the \textbf{UniTrans} have been tailored for optimal performance, the possibility persists that alternative prompt designs may exist with superior efficacy. Thus, we call for researchers to explore more about prompt design based on the three-step (i.e., Test Case Generation Phase, Translation Augmentation Phase, and Translation Repair Phase) framework we proposed, thereby advancing the state-of-the-art in automated code translation.

\begin{acks}
This work was partially supported by National Key R\&D Program under Grant No.2023YFB4503801, National Natural Science Foundation of China (Grant No. 62102233, 62302021, 62192731, 62192730, 62072007, 62192733, 61832009), Shandong Province Overseas Outstanding Youth Fund (Grant No. 2022HWYQ-043), Qilu Young Scholar Program of Shandong University, the General Research Fund (GRF) of the Research Grants Council of Hong Kong, the industry research funds of City University of Hong Kong (7005217,9220097,9220103,9229029,9229098,9678149), and the Key Program of Hubei under Grant JD2023008.
\end{acks}

\bibliographystyle{ACM-Reference-Format}
\bibliography{sample-base}


\end{document}